\documentclass[lettersize,journal]{IEEEtran}

\usepackage[utf8]{inputenc}
\usepackage[T1]{fontenc}
\usepackage[english]{babel}

\usepackage{amsmath, amssymb, amsfonts, amsthm}
\usepackage{thmtools, empheq}
\usepackage{nicefrac}

\usepackage{graphicx}
\usepackage{subcaption}
\usepackage{booktabs}
\usepackage{threeparttable, tabularx}
\usepackage{array}
\usepackage{caption}

\usepackage{enumitem}
\usepackage{changepage}
\usepackage{setspace}
\usepackage{float}
\usepackage{stfloats}
\usepackage{verbatim}

\usepackage{algorithm}
\usepackage{algpseudocode}

\usepackage{xcolor}
\usepackage{xparse}

\usepackage{url}
\usepackage{xr}
\usepackage{cite}
\usepackage{ifpdf}

\usepackage{lipsum}
\usepackage{textcomp}

\usepackage{hyperref}

\graphicspath{{./Figs/}}
\algnewcommand{\grayComment}[1]{\hfill \textcolor{gray}{\# #1}}

\declaretheorem[style=definition,qed=$\square$]{proposition}

\NewDocumentCommand{\multicite}{>{\SplitList{,}}m}{%
	\def\temp{}
	\ProcessList{#1}{\addcite}%
	\textup{(\temp)}
}

\newcommand{\addcite}[1]{%
	\ifx\temp\empty
	\xdef\temp{\ref{#1}}%
	\else
	\xdef\temp{\temp, \ref{#1}}%
	\fi
}

\begin{document}

\title{Subspace-Based Super-Resolution Sensing for Bi-Static ISAC with Clock Asynchronism}

\author{Jingbo~Zhao,~
	Zhaoming~Lu,~\IEEEmembership{Member,~IEEE,}
	J.~Andrew~Zhang,~\IEEEmembership{Senior Member,~IEEE,}
	\\
	Jiaxi Zhou,~
	Weicai Li,~ 
	and~Tao Gu,~\IEEEmembership{Fellow,~IEEE}
	
	{
	\IEEEcompsocitemizethanks{
		\IEEEcompsocthanksitem This work has been submitted to the IEEE for possible publication. Copyright may be transferred without notice, after which this version may no longer be accessible. 
		\IEEEcompsocthanksitem Jingbo Zhao, Zhaoming Lu, Jiaxi Zhou and Weicai Li are with the School of Information and Communication Engineering, Beijing University of Posts and Telecommunications, Beijing, China. (e-mail: \{zhjb, lzy0372, zhoujiaxi, liweicai\}@bupt.edu.cn).
		\IEEEcompsocthanksitem J. Andrew Zhang is with the School of Electrical and Data Engineering, University of Technology Sydney. (e-mail: andrew.zhang@uts.edu.au)
		\IEEEcompsocthanksitem Tao Gu is with the School of Computing, Macquarie University. (e-mail: tao.gu@mq.edu.au)
	}}
	\vspace{-15pt}
}

\maketitle

\begin{abstract}	
Bi-static sensing is an attractive configuration for integrated sensing and communications (ISAC) systems; however, clock asynchronism between widely separated transmitters and receivers introduces time-varying time offsets (TO) and phase offsets (PO), posing significant challenges. 
This paper introduces a signal-subspace-based framework that estimates decoupled angles, delays, and complex gain sequences (CGS) — the target-reflected signals — for multiple dynamic target paths. The proposed framework begins with a novel TO alignment algorithm, leveraging signal subspace or covariance, to mitigate TO variations across temporal snapshots, enabling coherent delay-domain analysis. Subsequently, subspace-based methods are developed to compensate for TO residuals and to perform joint angle-delay estimation. Finally, leveraging the high resolution in the joint angle-delay domain, the framework compensates for the PO and estimates the CGS for each target. The framework can be applied to both single-antenna and multi-antenna systems. 
Extensive simulations and experiments using commercial Wi-Fi devices demonstrate that the proposed framework significantly surpasses existing solutions in parameter estimation accuracy and delay resolution. Notably, it uniquely achieves a super-resolution in the delay domain, with a probability-of-resolution curve tightly approaching that in synchronized systems. 

\end{abstract}
\vspace{-1.3pt}

\vspace{-1.2pt}
\section{Introduction}
\IEEEPARstart{I}{ntegrated} sensing and communication (ISAC) technology has been identified as a key technology for future communications, e.g., 6G \cite{ITU2024} and Wi-Fi systems \cite{Survey_80211bf}. Leveraging widespread communication infrastructure, ISAC technology provides ubiquitous and seamless sensing services, which are integral to enabling smart applications \cite{Survey_ISAC}.

\begin{figure*}[t]
	\vspace{-18pt}
	\centering
	\includegraphics[width=17.2cm,height=4.65cm]{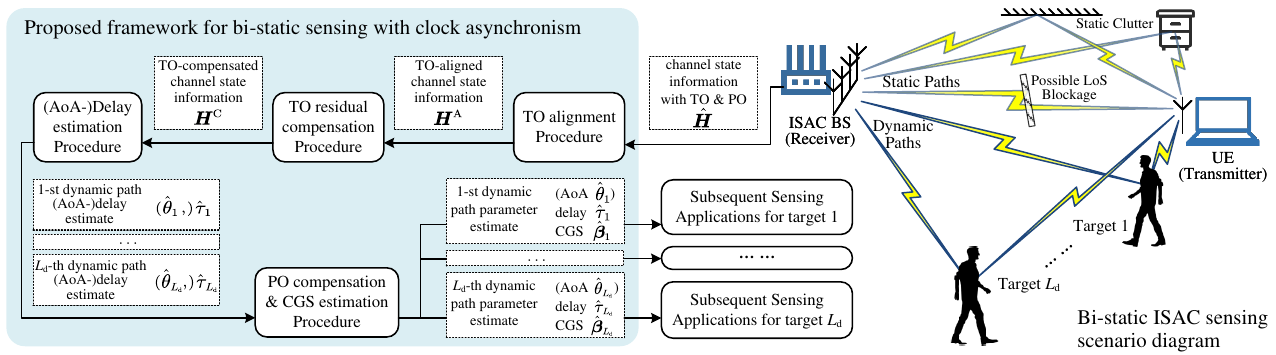}
	\vspace{-6pt}
	\caption{Diagram of bi-static sensing with asynchronous ISAC transceivers (right) and the proposed signal processing framework (left).}
	\label{fig:scenario_framework}
	\vspace{-15pt}
\end{figure*}

ISAC applications may require different setups. Among various setups, the bi-static sensing setup, as illustrated in \mbox{Fig. \ref{fig:scenario_framework},} stands out as the most attractive configuration \cite{Survey_bistatic}. 
This setup capitalizes on analyzing the multipath propagation patterns of communication signals sent by user equipment (UE) at the ISAC base station (BS) to estimate the state of targets, typically human bodies, enabling diverse applications such as device-free tracking \cite{Indotrack-track-CACC,CFSR-tracking-respiration, RSSI-tracking}, \mbox{activity recognition \cite{Widar3-gesture-CACC, sensor-gesture},} and vital signs monitoring \cite{FarSense, CFSR-tracking-respiration}. 
Importantly, the bi-static setup reuses existing communication infrastructure without altering the prevalent half-duplex architecture, thus maximizing the utilization efficiency of devices, spectrum, and power resources.

Clock asynchronism is a common problem in bi-static sensing, making it especially challenging in general environments characterized by multiple targets, static clutter, and non-line-of-sight (NLoS) conditions \cite{Survey_bistatic, Survey_challenges}. 
Given that the communication transmitter and receiver are typically separated in space and operate on their local clocks, the asynchronism between them introduces various distortions during the channel estimation, collectively presenting fast-varying time offset (TO) and phase offset (PO) in the sensing signals. 
Such offsets are generally tolerable in communications, where only short-term relative synchronization is required for each frame. However, for sensing applications, which typically requires phase-level absolute synchronization, these offsets impede coherent signal processing along subcarriers or temporal snapshots, invalidating the direct use of traditional delay and Doppler estimation and introducing ambiguity in the sensing results.
Moreover, in urban or indoor bi-static scenarios, multiple dynamic targets, the potential absence of a LoS path, and massive static clutter exacerbate the difficulty of addressing asynchronism offsets, and further impede the identification, resolution, and parameter estimation of dynamic target paths.

Several solutions have been developed to address clock asynchronism, supporting various bi-static sensing applications \cite{Widar3-gesture-CACC, Indotrack-track-CACC, FarSense, CFSR-tracking-respiration, Spotfi, Sharp, multisense, Optimal, Anchor, JUMP}. However, these solutions come with certain limitations.
Early works employ an \textit{offset cancellation} method in different domains. 
Specifically, solutions including cross-antenna cross-correlation \cite{Widar3-gesture-CACC, Indotrack-track-CACC}, cross-antenna signal ratio \cite{FarSense}, and cross-subcarrier signal ratio \cite{CFSR-tracking-respiration}, have been developed. 
These solutions calculate the conjugate multiplication or ratio of signals on different antennas/subcarriers to cancel out the PO, primarily aimed at recovering Doppler information in single-target scenarios. 
However, these approaches introduce nonlinear transformations or generate inter-modulation components, and perform poorly in multi-target scenarios. 
Recent studies \cite{Spotfi, Sharp, multisense, Optimal, Anchor, JUMP} resort to an \textit{offsets estimation and compensation} principle that leverages the consistency of TO and PO across propagation paths to approximately estimate and compensate for these offsets, thereby enabling coherent signal processing to estimate the relative delay and Doppler of targets. 
Specifically, research \cite{Spotfi, Sharp, multisense} estimate TO and PO by approximately extracting the line-of-sight (LoS) path or the dominant NLoS path as the reference to estimate TO and PO; research \cite{Optimal, Anchor, JUMP} utilize the overall response as the reference, where the static paths component is typically assumed to be dominant, to estimate the offsets. 
However, due to the limited bandwidth of communication systems, the reference components constructed in these solutions are still interfered with the dynamic target paths, leading to biases in TO and PO estimation relevant to the parameters of dynamic target paths. This can significantly degrade the target paths parameter estimation accuracy and the resolution, especially in NLoS and multi-target scenarios.

Distinct from existing solutions, this paper proposes a framework to address clock asynchronism by leveraging the following two key concepts, employed at different stages of the process: 1) the signal subspace or covariance, which captures the inherent structure of signals in the delay domain, and 2) a static paths component, clear from the dynamic paths, extracted via subspace methods. These concepts account for the impact of multiple dynamic target paths, enabling precise estimation and compensation of the TO and PO. 
Consequently, the proposed framework \textit{significantly enhances parameter estimation accuracy} and \textit{delay resolution}. 
Additionally, capitalizing on its high delay resolution, the framework \textit{allows for delay and Doppler estimation with a single-antenna system}. It also \textit{operates effectively in scenarios without a static LoS path} or \textit{where multiple dynamic paths may significantly contribute to the channel}, since both static and dynamic paths are exploited for addressing asynchronism. 

Building on these core concepts, the proposed framework is designed to broadly support bi-static sensing applications.  
Specifically, the framework uses asynchronous multiple-input multiple-output orthogonal frequency-division multiplexing (MIMO-OFDM) channel state information (CSI) to estimate the parameters of each dynamic target path, including the angle of arrival (AoA) (when a multi antenna receiver is employed), the delay, and the complex gain sequence (CGS). These parameters jointly provide a comprehensive description of the target paths, thereby enabling a wide range of sensing applications. Specially, the CGS parameter for each target can be understood as the target-reflection signal, with the influences of AoA, delay, asynchronism offsets, and interference of other dynamic targets decoupled, primarily encapsulating the target’s Doppler information. Compared to the traditional single-tone Doppler parameter, CGS offers a more accurate representation for targets exhibiting multiple Doppler components, such as a moving human body. 
Procedurally, the framework first aligns the TO across snapshots to enable coherent processing in the delay domain. It then compensates for residual TO and estimates the joint AoA-delay for dynamic target paths, and finally compensates for the PO and estimates the CGS for each target. Simulation evaluations and a practical use case on commercial Wi-Fi platforms demonstrate that our framework uniquely achieves super-resolution delay estimation and delivers precise CGS estimates for multiple targets, significantly outperforming existing techniques. 

The contributions of this paper are summarized as follows:
\begin{itemize}[leftmargin=*]	
	\item We propose a TO alignment algorithm for asynchronous OFDM CSI, which estimates and compensates for relative TO across snapshots, unlocking coherent processing across subcarriers and temporal snapshots. 
	Specifically, based on the maximum likelihood principle, we design a covariance-based and a subspace-based TO alignment method, catering respectively to stationary and non-stationary dynamic target signals. 
	Moreover, we improve computational efficiency by integrating the sparse fast Fourier transform (SFFT), making it particularly suitable for real-time processing.
	
	\item We develop a TO residual compensation scheme coupled with a modified MUSIC spectrum to accurately estimate the absolute delays of dynamic targets. 
	The compensation scheme utilizes a reference static paths response to restore the absolute delay information in the TO-aligned CSI. Concurrently, the modified MUSIC spectrum leverages this reference to suppress interference from static clutter, thereby enabling robust delay estimation for dynamic targets. 
	
	\item Based on the above algorithms, we provide a comprehensive sensing framework for bi-static ISAC with clock asynchronism by proposing supplementary algorithm for PO compensation and CGS estimation using the (joint AoA-)delay resolution. 
	This framework estimates the AoA, delay, and CGS (i.e., the decoupled target reflection signal) for multiple dynamic targets from asynchronous \mbox{(MIMO-)OFDM CSI,} providing universal support for sensing applications. 
	The framework applies to challenging scenarios, including those with single-antenna transceivers, or where multiple dynamic paths contribute to a significant power proportion, as well as scenarios with NLoS conditions and substantial static clutter.
	
	\item We conduct numerical simulations and practical experiments for device-free tracking using commercial Wi-Fi devices to evaluate and validate the proposed framework. The results demonstrate its significant advantages in parameter estimation accuracy and delay resolution over existing solutions, with only a modest and reasonable performance gap compared to a fully synchronized system. 
\end{itemize}

The rest of this paper is organized as follows. 
Sec. \ref{sec:system_model} presents the system and signal models and overviews the proposed framework. 
Sec. \ref{sec:TO_Alignment} details the TO alignment. 
Sec. \ref{sec:TO_compensation_delay_estimation} introduces the TO residual compensation and delay estimation. 
Sec. \ref{sec:CGS_estimation_extension} presents the CGS estimation and extends the framework to MIMO-ISAC. 
Sec. \ref{sec:evaluations} and \ref{sec:use_case} present the simulation and experiment results, respectively. 
Sec. \ref{sec:conclusion} concludes the paper. 

\vspace{3pt}
\begingroup
\small\setstretch{0.9}
\noindent\textbf{Notations:} $(\cdot)^{\intercal}$, $(\cdot)^{*}$, $(\cdot)^{\dagger}$, $(\cdot)^{-1}$, $(\cdot)^{+}$, $(\cdot)^{\frac{1}{2}}$ denote the transpose, conjugate, conjugate transpose, inverse, Moore–Penrose inverse, and square root of a matrix, respectively. 
$\odot$ and $\otimes$ denote the Hadamard product and Kronecker product.
$[\cdot]_{a}$ denotes the $a$-th entry of a vector or the $a$-th column of a matrix; $[\cdot]_{a:b}$ denotes the submatrix consisting  of columns $a$ to $b$ of a matrix, or the subvector containing entries $a$ to $b$ of a vector; $[\cdot]_{\substack{\!c:d \\ \!a:b}}$ denotes the submatrix formed by taking rows $c$ to $d$ and columns $a$ to $b$ of a matrix.
\hfill$\square$
\endgroup

\section{System Model}
\label{sec:system_model}
Consider a typical bi-static ISAC scenario in an indoor environment, as shown in Fig. \ref{fig:scenario_framework}. The setup comprises a UE and an ISAC BS. The ISAC BS measures the CSI with the UE, and utilizes the signal paths reflected from the nearby dynamic human targets for sensing applications. We presume that both the BS and UE remain static, which aligns with the practical usage of wireless routers as BS, and devices such as laptops or smart furniture equipped with Wi-Fi or cellular modules as UEs. 
This scenario is characterized by two primary attributes:
\begin{itemize}[leftmargin=*]
	\item \textit{Clock asynchronism}: Since the transmitter and receiver are spatially separated commercial communication devices, they operate on independent clocks, collectively inducing time-varying TO and PO in CSI measurements \cite{Survey_bistatic}.
	\item \textit{NLoS and static clutter}: The indoor environment is generally cluttered with numerous propagation paths reflected or scattered from the environmental objects, and the LoS path between the transmitter and receiver might be occluded.
\end{itemize} 

Next, we introduce the sensing signal model (i.e., the CSI measurement model) for this scenario, and analyze the challenges of coherent signal processing within this framework.

\subsection{The Signal Model}
For brevity, we consider a single-input single-output (SISO)-OFDM model. The proposed framework will be extended to MIMO-OFDM systems in Sec. \ref{sec:extension} by expanding the spatial dimension, realizing joint angle-delay estimation.
Assuming the system employs $K$ subcarriers, the frequency-response vector for a path with delay $\tau$ is given by
\vspace{-3pt}
\begin{equation}
	\boldsymbol{a}(\tau)=\mathrm{e}^{-\mathrm{j}2\pi\boldsymbol{f}\tau},
	\vspace{-3pt}
\end{equation}
where $\boldsymbol{f}={[0, \Delta f, .., (K-1)\Delta f]}^{\intercal}$ is the frequencies of the subcarriers, with $\Delta f$ denoting the subcarrier interval. 

Assuming that there are $L_\text{s}$ static propagation paths, including the potential LoS path and those reflected by static environmental objects, and $L_\text{d}$ dynamic propagation paths reflected by dynamic targets, the channel vector at time $\tilde{t}$ is given by $\boldsymbol{h}(\tilde{t}) \!\in\! \mathbb{C}^{K\!\times\! 1}$, 
\vspace{-3pt}
\begin{equation}
	\!\!\boldsymbol{h}(\tilde{t})\!=\!\bigg(\!\sum_{l=1}^{L_{\text{s}}}\!\beta_{\text{s},l}\boldsymbol{a}(\tau_{\text{s},l})\!+\!\!\sum_{l=1}^{L_\text{d}}\!  \beta_{l}\!(\tilde{t})\!\cdot\!\boldsymbol{a}\big(\!\tau_{l}(\tilde{t})\!\big)\!\bigg)\!\odot\!\Big(\!\mathrm{e}^{\mathrm{j}\varphi_{\text{o}}\!(\tilde{t})}\!\!\cdot\!\boldsymbol{a}\big(\!\tau_{\text{o}}(\tilde{t})\!\big)\!\Big)\!,\!\!
	\label{eq:channel_vector}
	\vspace{-3pt}
\end{equation}
where $\tau_{\text{s},l}$ and $\beta_{\text{s},l}$ denote the delay and complex gain of the $l$-th static path, respectively; $\tau_{l}(\tilde{t})$ and $\beta_{l}(\tilde{t})$ denote the delay and complex gain of the $l$-th dynamic path, respectively; $\tau_{\text{o}}(\tilde{t})$ and $\varphi_{\text{o}}(\tilde{t})$ are the TO and PO, respectively \cite{Survey_bistatic}. 

In a typical indoor bi-static ISAC scenario, the time-varying characteristics of the parameters in \eqref{eq:channel_vector} are as follows. 
The static path parameters, $\tau_{\text{s},l}$ and $\beta_{\text{s},l}$, remain constant unless there is a significant change in the static environment or movement of the transmitter or receiver. Therefore, these parameters are considered time-invariant in this paper. 
The dynamic paths delay $\tau_{l}(\tilde{t})$ (and AoA) typically has seconds-level coherence times, denoted as $\text{T}_{\text{CP}}$. This reflects the typical movement speeds of human targets, whose displacements over several seconds are generally much smaller than the system's delay (and angular) resolution. 
Conversely, the dynamic paths' complex gains $\beta_{l}(\tilde{t})$ are rapidly time-varying, as minor wavelength-level displacements in dynamic target body parts can cause significant phase changes in $\beta_{l}(\tilde{t})$. 
Furthermore, the TO $\tau_{\text{o}}(\tilde{t})$ and PO $\varphi_{\text{o}}(\tilde{t})$, are also rapidly time-varying due to discontinuous transmission, changes in synchronization point, and fast-varying compensation residual \cite{Survey_bistatic}. This paper considers the most challenging scenario where $\beta_{l}(\tilde{t})$, $\tau_{\text{o}}(\tilde{t})$ and $\varphi_{\text{o}}(\tilde{t})$ may vary across channel snapshots. 

Leveraging the time-varying characteristics of the channel parameters, we formulate a CSI measurement model within a coherent processing interval (CPI) for dynamic paths delay and AoA. 
Consider a $q$-th CPI starting at time $t_0^q$, containing $T$ consecutive CSI snapshots with the interval of $\Delta t$, where $T \!\cdot\! \Delta t \!<\! \text{T}_{\text{CP}}$ is satisfied to ensure that the dynamic path delays are constant in the $T$ snapshots. 
We first rewrite the static paths component in \eqref{eq:channel_vector} as a vector \mbox{$\boldsymbol{h}_{\text{s}}\!\!=\!\!\sum_{l=1}^{L_\text{s}}\!\beta_{\text{s},l}\boldsymbol{a}\!(\tau_{\text{s},l}) \!\in\!\! \mathbb{C}^{\!K\!\times\!1}\!\!$, which} is constant across snapshots and CPIs. \mbox{This vector form cor-} responds to a ``merged static path'' as all static paths are linearly dependent across snapshots \cite{spatial_smoothing}, ensuring the source covariance matrix to be full-rank, and it adopts the cluttered multipath environment, where static path number can be large. 
Then, we denote the delay of the $l$-th dynamic path during the $q$-th CPI as $\tau_{l}^{q}$ by omitting its variation within a CPI according to the aforementioned time-varying characteristic. 
The $t$-th CSI snapshot within the $q$-th CPI is given by $\hat{\boldsymbol{h}}^{q}_{t} \!\in\! \mathbb{C}^{K\!\times\!1}$,
\vspace{-6pt}
\begin{equation}
	\hat{\boldsymbol{h}}^{q}_{t}=\bigg(\boldsymbol{h}_{\text{s}}+\sum_{l=1}^{L_\text{d}} \beta_{l,t}^{q} \cdot \boldsymbol{a}(\tau_{l}^{q}) \bigg)\odot\Big(\mathrm{e}^{\mathrm{j}\varphi^{q}_{\text{o},t}}\cdot\boldsymbol{a}(\tau^{q}_{\text{o},t})\Big)+\boldsymbol{z},
	\label{eq:CSI_snapshot}
	\vspace{-5pt}
\end{equation}
where $\tau^{q}_{\!\text{o},t}\!\!=\!\!\tau_{\!\text{o}}\!(t_0^q\!+\!t\!\cdot\! \Delta t)$, $\varphi^{q}_{\!\text{o},t}\!\!=\!\!\varphi_{\!\text{o}}\!(t_0^q\!+\!t\!\cdot\! \Delta t)$, and $\beta_{\!l,t}^{q}\!\!=\!\!\beta_{l}\!(t_0^q\!+\!t\!\cdot\! \Delta t)$ are the TO, PO, and the $l$-th dynamic path's complex gain at the time of the snapshot, respectively; and $\boldsymbol{z}$ is the additive complex Gaussian noise.
The reasons that we consider time-varying complex path gains rather than conventional models based on fixed path gains and Doppler frequencies are two-fold, as detailed in \cite{Performance_Bound_Asyn}. First, the path gains may indeed be time-varying; and second, some sensing applications, such as activity recognition, are better based on complex gains.

By stacking the snapshots, the CSI measurements within the $q$-th CPI can be represented as the matrix $\hat{\boldsymbol{H}}^{q}\!\in\!\mathbb{C}^{K\!\times\! T}$, 
\vspace{-3pt}
\begin{equation}
	\hat{\boldsymbol{H}}^{q} = \big(\boldsymbol{A}^{q} \cdot {\boldsymbol{B}^{q}}^{\intercal}\big) \odot \boldsymbol{\Phi}(\boldsymbol{\tau}^{q}_{\text{o}}, \boldsymbol{\varphi}^{q}_{\text{o}}) + \boldsymbol{z},
	\label{eq:CSI_measurements}
	\vspace{-3pt}
\end{equation}
where $\boldsymbol{A}^{\!q}\!\!=\!\!\big[\boldsymbol{h}_{\text{s}}\, \boldsymbol{A}^{\!q}_{\!\text{d}}\big] \!\!\in\!\! \mathbb{C}^{\!K\!\times\! (\!L_{\text{d}}+1\!)}$ is the frequency response matrix,  with $\boldsymbol{A}^{\!q}_{\!\text{d}}\!\!=\!\!\big[\boldsymbol{a}\!(\!\tau^{q}_{1}),..,\boldsymbol{a}\!(\!\tau^{q}_{\!L_{\text{d}}}\!)\big]\!$ denoting the frequency response submatrix corresponding to the dynamic paths; 
\mbox{$\!\boldsymbol{B}^{q} \!\!=\!\! \big[\boldsymbol{1}_{\!T\!\times\! 1}\, \boldsymbol{B}^{q}_{\text{d}}\big] \!\!\in\!\! \mathbb{C}^{T\!\times\! (\!L_{\text{d}}+1\!)}$} is the signal matrix, with $\boldsymbol{B}^{q}_{\text{d}}\!\!=\!\!\big[\boldsymbol{\beta}^{q}_{1},..,\boldsymbol{\beta}^{q}_{L_{\text{d}}}\big]$ denoting the signal submatrix corresponding to the dynamic paths, and $\boldsymbol{\beta}^{q}_{l} \!\!=\!\! \big[\beta^{q}_{\!l\!,1}\!,..,\beta^{q}_{\!l,\!T}\big]^{\!\intercal}\!\!$ is the CGS of the \mbox{$l$-th} dynamic path;  
$\boldsymbol{\Phi}\!(\!\boldsymbol{\tau}^{\!q}_{\!\text{o}}, \boldsymbol{\varphi}^{\!q}_{\!\text{o}})\!\!=\!\!\big[{\mathrm e}^{{\mathrm j}\varphi^{\!q}_{\!\text{o},\!1}}\boldsymbol{a}\!(\tau^{\!q}_{\!\text{o},\!1}\!),\!.., {\mathrm e}^{\mathrm{j}\varphi^{\!q}_{\!\text{o},\!T}}\boldsymbol{a}\!(\tau^{\!q}_{\!\text{o},\!T}\!)\big]$ is the TO and PO induced \mbox{phase offset,} 
with snapshot-unique TO $\!\boldsymbol{\tau}^{\!q}_{\!\text{o}} \!\!=\!\! \big[\tau^{\!q}_{\!\text{o},\!1},\!..,\! \tau^{\!q}_{\!\text{o},\!T}\big]^{\!\intercal}\!\!\!$ and PO $\!\boldsymbol{\varphi}^{\!q}_{\text{o}} \!\!=\!\! \big[\varphi^{\!q}_{\!\text{o},\!1},\!..,\! \varphi^{\!q}_{\!\text{o},\!T}\big]^{\!\intercal}\!\!\!$.

\subsection{Signal Subspace Preliminary} 
\label{sec:subspace_preliminary}
\begin{figure}[t]
	\vspace{-12pt}
	\includegraphics[width=9cm, height=4.0cm]{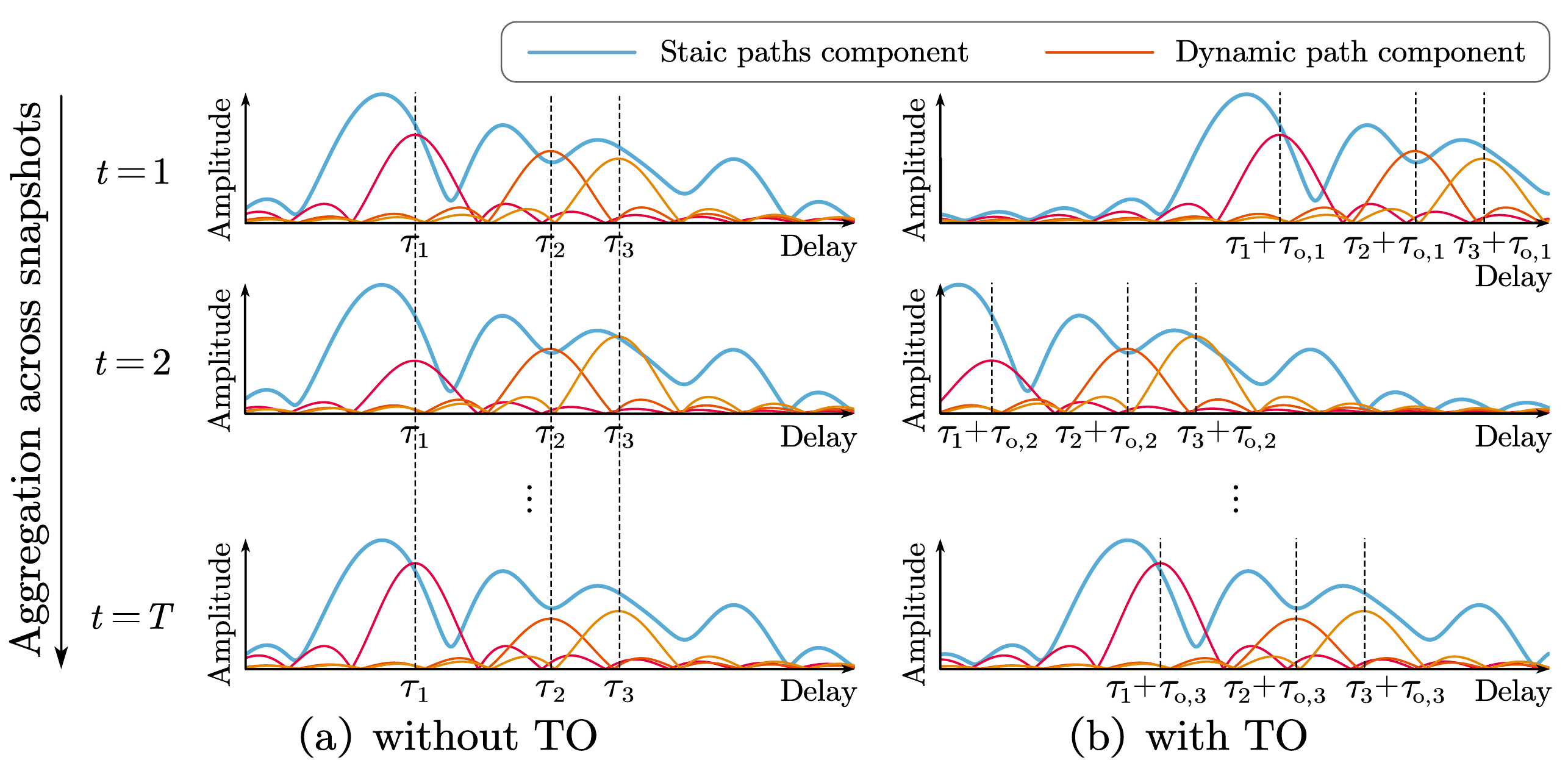}
	\vspace{-17pt}
	\caption{Impact of time-varying TO on coherent processing across temporal snapshots.}
	\label{Fig_PDP}
	\vspace{-3pt}
\end{figure}

The signal subspace $\mathcal{U}_\mathrm{S}$ is defined as the subspace spanned by the frequency response vectors of linearly independent signal paths within the sensor array space $\mathbb{C}^{K}$. 
Focusing on a specific CPI and omitting the CPI index $q$, the signal subspace in our model is given by $\mathcal{U}_\mathrm{S}\!=\!\text{span}\{\boldsymbol{h}_\mathrm{s}, \boldsymbol{a}(\tau_1),\!.., \boldsymbol{a}(\tau_{L_\mathrm{d}}\!)\!\}\!
=\!\big\{\!\boldsymbol{v}\!\!\in\!\!\mathbb{C}^{\!K\!\times1}|\boldsymbol{v}\!\!=\!\!\boldsymbol{A}\!\!\cdot\!\boldsymbol{c}, \boldsymbol{c}\!\!\in\!\!\mathbb{C}^{(L_{\text{d}}\!+\!1)\!\times\!1}\!\big\}$, and the noise subspace is defined as the orthogonal complement of the signal subspace, $\mathcal{U}_\mathrm{N}=\mathcal{U}_\mathrm{S}^{\perp}$. 

In synchronized systems, the signal subspace can be estimated by the eigenvalue decomposition (EVD) of the sensor covariance matrix. 
Denote the CSI matrix of the synchronous system as $\hat{\boldsymbol{H}}^{\!\text{syn}}\!\!$, which corresponds to $\hat{\boldsymbol{H}}^{\!q}$ in \eqref{eq:CSI_measurements}, \mbox{with the asyn-} chronism term $\!\boldsymbol{\Phi}\!(\!\boldsymbol{\tau}^{q}_{\!\text{o}}\!, \boldsymbol{\varphi}^{q}_{\!\text{o}}\!)\!$ replaced by an all-one matrix.
The corresponding sensor covariance matrix, hereafter referred to as covariance matrix, over a large snapshot number $T\!$, is given by
\vspace{-5pt}
\begin{equation}
	\vspace{-4pt}
	\widehat{\boldsymbol{R}}_{\mathrm{H}}^{\text{syn}}
	=\frac{1}{T} \hat{\boldsymbol{H}}^{\text{syn}} {\,\!\hat{\boldsymbol{H}}^{\text{syn}}}^{\dagger}
	\xrightarrow{T \to \infty}
	\underbrace{\boldsymbol{A}\boldsymbol{R}_{\mathrm{b}}\boldsymbol{A}^{\dagger}}_{\boldsymbol{R}_{\text{S}}}
	+\underbrace{\sigma^2\boldsymbol{\mathrm{I}}}_{\boldsymbol{R}_{\text{N}}},
	\label{eq:cross-frequency-correlation}
	\vspace{-4pt}
\end{equation}
where $\boldsymbol{R}_{\mathrm{b}}\!\!\in\!\!\mathbb{C}^{(L_{\text{d}}\!+\!1)\!\times\!(L_{\text{d}}\!+\!1)\!}$ is the full-rank source covariance matrix, $\boldsymbol{R}_{\text{S}}$ is the signal covariance matrix of rank $L_{\text{d}}\!+\!1$, and $\boldsymbol{R}_{\text{N}}$ is the noise covariance matrix. 
Represent $\boldsymbol{R}_{\text{S}}$ in the form of EVD with descending eigenvalues, $\boldsymbol{R}_{\text{S}}\!=\!\boldsymbol{U}\boldsymbol{\Lambda}_{\text{S}}\boldsymbol{U}^{\dagger}$, 
where $\boldsymbol{\Lambda}_{\text{S}}\!=\!\text{diag}\!\big([\lambda_1,..,\lambda_{L_{\text{d}}+1}, 0,..,0]\big)$ contains the $L_{\text{d}}\!+\!1$ non-zero eigenvalues and $K\!-\!L_{\text{d}}\!-\!1$ zero eigenvalues, 
and \mbox{$\boldsymbol{U}\!\!=\!\!\big[ \boldsymbol{U}_{\!\text{S}}\,\boldsymbol{U}_{\!\text{N}}\big]$,} with $\boldsymbol{U}_{\!\text{S}}\!\!=\!\!\big[\boldsymbol{u}_1,\!.., \boldsymbol{u}_{L_{\text{d}}\!+\!1}\big]$ and $\boldsymbol{U}_{\!\text{N}}\!\!=\!\!\big[\boldsymbol{u}_{L_{\text{d}}\!+\!2},\!.., \boldsymbol{u}_K\big]$ composing of the eigenvectors corresponding to the non-zero eigenvalues and the zero eigenvalues, respectively.
Each eigenvector corresponding to a non-zero eigenvalues, $\boldsymbol{u}_i, i \!\in\! [1, L_{\text{d}}\!+\!1]$, satisfies $\boldsymbol{A}\boldsymbol{R}_{\text{b}}\boldsymbol{A}^{\!\dagger}\boldsymbol{u}_i \!\!=\!\! \lambda_i\boldsymbol{u}_i $, thus belongs to $\mathcal{U}_\mathrm{S}$, $\boldsymbol{u}_i \!\!=\!\! \boldsymbol{A}\!\cdot\!\big(\!\frac{1}{\lambda_i}\boldsymbol{R}_{\text{b}} \boldsymbol{A}^{\!\dagger}\boldsymbol{u}_i\!\big)\!\!\in\!\mathcal{U}_\mathrm{S}$. 
Therefore, $\boldsymbol{U}_{\!\text{S}}$ serves as an orthonormal generator matrix of the signal subspace $\mathcal{U}_\mathrm{S}$. 
In practice, the receiver acquires the corresponding sensor covariance matrix $\widehat{\boldsymbol{R}}_{\mathrm{H}}^{\text{syn}}$, which contains the noise covariance matrix. 
Levarging the structure of $\boldsymbol{R}_{\text{N}}$, $\widehat{\boldsymbol{R}}_{\mathrm{H}}^{\text{syn}}\!\boldsymbol{u}_i \!=\!\boldsymbol{A}\boldsymbol{R}_{\text{S}}\boldsymbol{A}^{\!\dagger}\boldsymbol{u}_i \!+\! \sigma^2\boldsymbol{u}_i \!=\! (\lambda_i\!+\!\sigma^2)\boldsymbol{u}_i\!\!$, implying that the EVD of $\widehat{\boldsymbol{R}}_{\mathrm{H}}^{\text{syn}}$ can be expressed as $\widehat{\boldsymbol{R}}_{\mathrm{H}}^{\text{syn}}\!=\!\boldsymbol{U}\!\boldsymbol{\Lambda}_{\text{H}}\boldsymbol{U}^{\dagger}\!$, where $\boldsymbol{\Lambda}_{\text{H}}\!=\!\text{diag}\!\big([\lambda_1\!+\!\sigma^2,..,\lambda_{L_{\text{d}}+1}\!+\!\sigma^2, \sigma^2,..,\sigma^2]\big)$. 
Consequently, the orthonormal generator matrix of the signal subspace, $\boldsymbol{U}_{\!\text{S}}$, can be estimated as the eigenvectors corresponding to the $L_{\text{d}}+1$ largest eigenvalues of $\widehat{\boldsymbol{R}}_{\mathrm{H}}^{\text{syn}}$. 

The obstacles to applying the subspace signal processing for path parameter estimation with clock asynchronism and static clutter are three-fold. 
Firstly, \textit{the signal subspace is hard to estimate due to variance across snapshots of the TO}. Intuitively, the estimation of the signal subspace via \eqref{eq:cross-frequency-correlation} in synchronized systems relies on snapshots stacking, i.e., aggregation of delay-domain information across snapshots, as shown in Fig. \ref{Fig_PDP}.a. However, in the asynchronous system, the snapshots are shifted differently by the time-varying TO, as shown in Fig. \ref{Fig_PDP}.b. 
Secondly, the unknown time shift by the TO obscures the true propagation delay of the signal, resulting in the \textit{loss of absolute delay information}. 
Similarly, signal processing in the Doppler domain also faces these challenges due to the time-varying PO. 
Thirdly, the signal subspace in our problem contains the additional base vector $\boldsymbol{h}_\mathrm{s}$ for the merged static paths, introducing uncertainty on the peaks in the classic MUSIC spectrum, \textit{interfering with the detection of the targets}.

\subsection{Signal Processing Framework}

This work presents a subspace-based signal processing framework for bi-static sensing using asynchronous transceivers in challenging scenarios with multiple dynamic targets, NLoS conditions and static clutter, \mbox{as shown in Fig. \ref{fig:scenario_framework}.} 
\begin{itemize}[leftmargin=*]
	\item Firstly, the asynchronous CSI snapshots $\hat{\boldsymbol{H}}$, containing time-varying TO $\boldsymbol{\tau}_\text{o}$ and PO $\boldsymbol{\varphi}_\text{o}$, are processed by a TO alignment procedure, which estimates and compensates for the relative TO across snapshots within a CPI. The generated aligned snapshots $\boldsymbol{H}^{\text{A}}$ has the TO $\boldsymbol{\tau}_\text{o}$ reduced into a constant TO residual $\tau^{\text{r}}$, enabling the application of conventional coherent processing algorithms, yet the path delays are still skewed by $\tau^{\text{r}}$. (Sec. \ref{sec:TO_Alignment})
	\item Subsequently, a TO residual compensation procedure is applied to $\!\boldsymbol{H}^{\text{A}}\!$ to eliminate $\tau^{\text{r}}\!$, generating the compensated snapshots $\boldsymbol{H}^{\text{C}}\!$, which is free of TO but still contains \mbox{PO $\boldsymbol{\varphi}_{\!\text{o}}$}. Using $\boldsymbol{H}^{\text{C}}$, the delays of the dynamic paths, $\widehat{\tau}_{1},\!..,\widehat{\tau}_{L_{\text{d}}}$, are estimated through a modified MUSIC spectrum, where the interference from static clutter is suppressed. (Sec. \ref{sec:TO_compensation_delay_estimation})
	\item Lastly, a CGS estimation algorithm utilizes the delay estimates $\widehat{\tau}_{1},\!..,\widehat{\tau}_{L_{\text{d}}}$ to separate the multipath signals from the TO-compensated snapshots $\boldsymbol{H}^{\text{C}}\!$, thereby compensating for the PO $\boldsymbol{\varphi}_\text{o}$ and estimating the CGS, $\hat{\boldsymbol{\beta}}_{l}$, for each dynamic path. (Sec. \ref{sec:CGS_estimation})
\end{itemize}
Additionally, the framework is extended to the multi-antenna model to incorporate AoA estimation and to utilize joint spatial-frequency diversity and joint angle-delay resolution, which is presented in Sec. \ref{sec:extension}. Note that a superscript $q$ may be added to these variables to indicate that they are specific to the $q$-th CPI.

\section{Relative TO Alignment}
\label{sec:TO_Alignment}

\begin{figure*}[t]
	\vspace{-18pt}
	\centering
	\includegraphics[width=17.1cm, height=4.8cm]{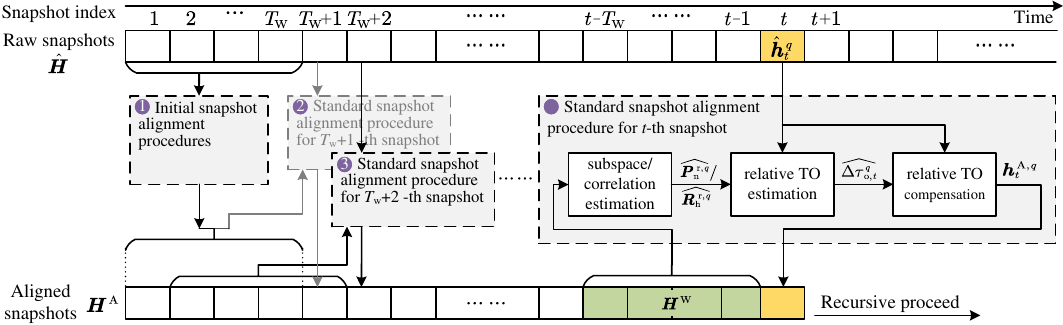}
	\vspace{-5pt}
	\caption{Diagram of the recursive TO alignment algorithm.}
	\label{Fig_alignment}
	\vspace{-15pt}
\end{figure*}

We propose a recursive algorithm to align the time-varying TO across the CSI snapshots. 
This algorithm processes the CSI measurements, $\hat{\boldsymbol{H}}$, which includes the snapshot-specific TO $\boldsymbol{\tau}_{\!\text{o}}$. It estimates and compensates for the relative TO between snapshots, yielding an aligned CSI matrix, $\boldsymbol{H}^{\text{A}}$. The algorithm ensures that the TO in $\boldsymbol{H}^{\text{A}}\!$ is consistently aligned within a CPI. For example, in any CPI segment of $\boldsymbol{H}^{\text{A}}\!$, specifically the one indexed by $q$, $\boldsymbol{H}^{\text{A}, q}\!$, the TO reduces to a residual $\tau_{\text{o}}^{\mathrm{r},q}\!$ that is common across all snapshots within the CPI. Next, we first provide an overview of the alignment workflow, and then provide detailed descriptions of each procedure. 

The workflow of the TO alignment algorithm is illustrated in Fig. \ref{Fig_alignment}, and listed as follows:
\begin{itemize}[leftmargin=*]
	\item The algorithm first performs an \textit{initial snapshots alignment procedure}, to align the first $T_{\text{w}}$ snapshots  ($T_{\text{w}}\cdot \Delta t < \text{T}_{\text{CP}}$), indexed by $t\!\in\![1,\ldots, T_{\text{w}}]$. 
	\item Then, the algorithm performs a \textit{standard snapshot alignment procedure} for each subsequent incoming snapshot, indexed by $t\!\!\in\!\![T_{\text{w}}{\scalebox{1.0}+}1\!,\!..,\!\infty]$. Use the $t$-th snapshot within the $q$-indexed CPI, $\hat{\boldsymbol{h}}_{t}^{q}$, as an example. The standard snapshot alignment procedure estimates the signal subspace or covariance from a sliding window $\boldsymbol{H}^{\text{w}}$ containing the previous $T_{\text{w}}$ aligned snapshots, and estimates and compensates for the relative TO in the incoming snapshot with respect to the previously aligned snapshots, obtaining the aligned new snapshot $\boldsymbol{h}^{\text{A},q}_{t}$. 
	\item Next, the sliding window $\boldsymbol{H}^{\text{w}}$ is updated using $\boldsymbol{h}^{\text{A},q}_{t}$, and the standard snapshot alignment procedure for the following snapshots recursively proceeds. 
\end{itemize}

After the application of the snapshots alignment algorithm, the $q$-th CPI of the resulting CSI matrix, $\boldsymbol{H}^{\text{A},q}$, ignoring the alignment error, can be represented as
\vspace{-3pt}
\begin{equation}
	\boldsymbol{H}^{\text{A},q} = \text{diag}\big(\boldsymbol{a}(\tau_{\text{o}}^{\mathrm{r},q})\big) \cdot \boldsymbol{A}^{q} \cdot {\boldsymbol{B}^{q}}^{\intercal} \cdot \text{diag}\big(\boldsymbol{\varphi}^{q}_{\text{o}}\big) + \boldsymbol{z}.
	\label{eq:H_aligned}
	\vspace{-3pt}
\end{equation}
With the TO aligned, $\!\boldsymbol{H}^{\text{A},q}\!$ is suitable for snapshot-stacking techniques such as MUSIC algorithm, although the resulting delay estimates remain shifted by a TO residual $\tau_{\text{o}}^{\mathrm{r},q}\!$. Note that the TO residual $\!\tau_{\text{o}}^{\mathrm{r},q}\!$ may vary across different CPIs due to cumulative alignment errors, stemming from changes in the dynamic paths delay and thus the \mbox{signal subspace or covariance.}

For the \textit{standard snapshot alignment procedure}, we propose two methods: subspace-based and covariance-based. The design and derivation of these methods are detailed in Sec. \ref{sec_snapshot_alignment_subspace} and Sec. \ref{sec_snapshot_alignment_covariance}, respectively. These approaches target distinct characteristics of path gain parameters. Specifically, the covariance-based method assumes that the path gains are statistically wide-sense stationary, whereas the subspace-based one does not require any prior assumptions about the path gain distribution. Theoretically, the covariance-based method is expected to yield higher accuracy in scenarios where all dynamic path gains are stationary. Conversely, the subspace-based method is more effective for general non-stationary path gain parameters, such as those exhibiting target glittering, i.e., drastic changes in target path powers, which can lead to sharp alterations in signal covariance.

The \textit{initial snapshots alignment procedure} is based on the standard procedure, and hence its design and derivation are detailed in Sec. \ref{sec_initial_TO_alignment}. 
We further \textit{accelerate the algorithm} by integrating SFFT to reduce the computational complexity of the snapshot alignment procedure. This acceleration, combined with our recursive framework, makes the algorithm particularly effective for real-time processing. Detailed discussions on algorithm acceleration, along with the complexity analysis and a summary of the algorithm, are presented in Sec. \ref{sec_TO_alignment_acceleration}.

\subsection{Subspace-Based Snapshot Alignment Procedure}
\label{sec_snapshot_alignment_subspace}

Consider aligning a $t$-th channel snapshot within the $q$-th CPI, $\hat{\boldsymbol{h}}_t^q$, expressed by \eqref{eq:CSI_snapshot}. 
At this point, the sliding window, $\boldsymbol{H}^{\text{w}}$, which contains the $T_{\text{w}}$ previously aligned snapshots, is given by a submatrix of \eqref{eq:H_aligned},  $\boldsymbol{H}^{\text{w}} = \big[\boldsymbol{H}^{\text{A},q}\big]_{\!t\!-\!T_{\!\text{w}}:t\!-\!1}$. 
Rewrite $\hat{\boldsymbol{h}}_t^q$ into matrix form, 
\vspace{-3pt}
\begin{equation}
	\hat{\boldsymbol{h}}_t^q = \text{diag}\big(\boldsymbol{a}(\Delta \tau_{\text{o},t}^{q})\big) \cdot \boldsymbol{A}^{\text{r},q} \cdot {\boldsymbol{b}^{q}_t} \cdot \mathrm{e}^{\mathrm{j}\varphi_{\text{o},t}^{q}} + \boldsymbol{z}.
	\label{eq:snapshot}
	\vspace{-3pt}
\end{equation}
where $\boldsymbol{b}^{q}_t = \big[1, \beta^{q}_{1,t},\ldots,\beta^{q}_{L_{\mathrm{d}},t}\big]^{\intercal}$ represent the complex gains of all paths (including the merged static path) corresponding to the $t$-th snapshot;
\vspace{-3pt}
\begin{equation}
	\boldsymbol{A}^{\text{r},q} = \text{diag}\big(\boldsymbol{a}(\tau_{\text{o}}^{\text{r},q})\big) \cdot \boldsymbol{A}^{q}
	\label{eq:A_shifted}
	\vspace{-3pt}
\end{equation}
is the frequency response matrix shifted by the TO residual; and $\Delta \tau_{\text{o},t}^{q} = \tau_{\text{o},t}^{q} - \tau_{\text{o}}^{\text{r},q}$ is the relative TO, with $\tau_{\text{o},t}^{q}$ and $\tau_{\text{o}}^{\text{r},q}$ being the TO in the $q$-th snapshot and the TO residual in $\boldsymbol{H}^{\text{w}}$, respectively. 
The snapshot alignment procedure is to estimate and compensate for the relative TO $\Delta \tau_{\text{o},t}^{q}$, thus aligning the snapshot to the sliding window.

\begin{proposition}[Subspace-Based Relative TO Estimation]
	\label{prop:TO_estimation_subspace}
	With additive complex Gaussian noise, the maximum likelihood estimate (MLE) of the relative TO at the $t$-th snapshot, $\widehat{\Delta \tau_{\text{o},t}^{q}}$, is given by
	\vspace{-3pt}
	\begin{equation}
		\setlength{\jot}{-6pt}
		\!\!\widehat{\Delta \tau_{\text{o},t}^{q}} = \arg\!\min_{\widetilde{\Delta\tau_{\text{o},t}^{q}}} \,
		{\hat{\boldsymbol{h}}_t^q\,\!}^{\dagger} \!\!\cdot\! \text{diag}\Big(\!\boldsymbol{a}\big(\!\widetilde{\Delta\tau_{\text{o},t}^{q}}\!\big)\!\Big) \!\cdot\! \boldsymbol{P}_{\mathrm{n}}^{\,\text{r},q} \cdot \text{diag}\Big(\boldsymbol{a}^{\!*}\!\big(\widetilde{\Delta\tau_{\text{o},t}^{q}}\big)\!\Big) \!\cdot\! {\hat{\boldsymbol{h}}_t^q} \!,
		\label{eq:TO_estmation_subspace}
		\vspace{-9pt}
	\end{equation}
	where 
	\vspace{-3pt}
	\begin{equation}
		\boldsymbol{P}_{\mathrm{n}}^{\,\text{r},q} \triangleq \boldsymbol{I}\!-\!\boldsymbol{A}^{\text{r},q}\big(\!{\boldsymbol{A}^{\text{r},q}}^{\dagger}\!\boldsymbol{A}^{\text{r},q}\big)^{\!\!-\!1}\!{\boldsymbol{A}^{\text{r},q}}^{\dagger}
		\label{eq:projection_shifted}
		\vspace{-3pt}
	\end{equation}
	is the \textit{projection matrix} into the noise subspace, shifted by the TO residual. The projection matrix $\boldsymbol{P}_{\mathrm{n}}^{\,\text{r},q}$ is treated as a known parameter, which is obtained using the previously aligned snapshots $\boldsymbol{H}^{\text{w}}$. \newline
	\noindent\textit{Derivation:} see Appendix \ref{Apd:TO_estimation_subspace}. 
\end{proposition}

Intuitively, Proposition \ref{prop:TO_estimation_subspace} shows that the MLE of relative TO is equivalent to a subspace method that minimizes the projection of the snapshot into the TO-residual-shifted noise subspace. This indicates that the MLE relies on the structure of the signal subspace, since the distribution of the path gain parameters is unknown. 

Next, we introduce the acquisition of the projection matrix $\boldsymbol{P}_{\mathrm{n}}^{\,\text{r},q}$. 
The estimate of the shifted projection matrix $\boldsymbol{P}_{\mathrm{n}}^{\,\text{r},q}$ is generated using $\boldsymbol{H}^{\text{w}}$ via the signal subspace estimation method presented in Sec. \ref{sec:subspace_preliminary}. Specifically, the covariance matrix, shifted by the TO residual $\tau_{\text{o}}^{\text{r}, q}$, is estimated as
\vspace{-3pt}
\begin{equation}
	\boldsymbol{R}_{\text{w}} = \boldsymbol{H}^{\text{w}}{\boldsymbol{H}^{\text{w}}}^{\dagger}. 
	\label{eq:covariance}
	\vspace{-3pt}
\end{equation}
Applying eigenvalue decomposition to the covariance matrix generates
\vspace{-3pt}
\begin{equation}
	\boldsymbol{R}_{\text{w}}=\boldsymbol{U}\boldsymbol{\Lambda}\boldsymbol{U}^{\dagger},
	\label{eq:EVD}
	\vspace{-4pt}
\end{equation}
where the columns of $\boldsymbol{U}\!=\!\big[\boldsymbol{u}_1,\ldots,\boldsymbol{u}_K\big]$ are orthonormal eigenvectors, and $\boldsymbol{\Lambda}\!=\!\text{diag}\big([{\lambda}_1,\ldots,{\lambda}_K]\big)$ is a diagonal matrix formed by eigenvalues in descending order. Following the derivation in Sec. \ref{sec:subspace_preliminary}, the eigenvectors corresponding to the $d$ largest eigenvalues form the orthonormal basis of the signal subspace, where $d$ is the signal subspace dimension. Thus, the projection matrix into the noise subspace, $P_{\mathrm{n}}^{\,\text{r},q}$, is estimated as
\vspace{-6pt}
\begin{equation}
	\widehat{\boldsymbol{P}_{\mathrm{n}}^{\,\text{r},q}}=\sum_{i=\hat{d}+1}^{K} \boldsymbol{u}_i \boldsymbol{u}_i^{\dagger},
	\label{eq:Pn}
	\vspace{-6pt}
\end{equation}
where $\hat{d}$ is the estimate of the signal subspace dimension, and can be determined using the minimum description length principle \cite{MDL}, 
\vspace{-5pt}
\begin{equation}
	\setlength{\jot}{-8pt}
	\begin{aligned}
		&\hat{d}=\arg\!\min_d \,F_{\text{MDL}}(d), \,  \\
		&\text{where}\,\, F_{\text{MDL}}(d)\!=\!T_{\text{w}}(K\!-\!d)\ln\!\!\Bigg(\!\frac{\frac{1}{K\!-\!d} \sum_{i=1+d}^{K}\hat{\lambda}_i}{\big(\!\prod_{i=1+d}^{K}\hat{\lambda}_i\big)^{\!\frac{1}{K\!-\!d}}}\!\Bigg).
	\end{aligned}
	\label{eq:MDL}
	\vspace{-4pt}
\end{equation}

Then, the relative TO at the $t$-th snapshot, $\Delta \tau_{\text{o},t}^{q}$, can be estimated using \eqref{eq:TO_estmation_subspace}, by substituting 
$\boldsymbol{P}_{\mathrm{n}}^{\,\text{r},q}$ with the estimate $\widehat{\boldsymbol{P}_{\mathrm{n}}^{\,\text{r},q}}$, and the $t$-th snapshot is aligned as 
\vspace{-4pt}
\begin{equation}
	\boldsymbol{h}_{t}^{\text{A},\,q} \leftarrow \text{diag}\big(\boldsymbol{a}^{\!*}\!(\widehat{\Delta\tau_{\text{o},t}^{q}})\!\big) \!\cdot\! {\hat{\boldsymbol{h}}_t^q} . 
	\label{eq:snapshot_alignment}
	\vspace{-2pt}
\end{equation}

The major computational complexity of the subspace-based standard snapshot alignment procedure includes ${O}(K^2)$ for updating the covariance matrix $\boldsymbol{R}_{\text{w}}$, ${O}(K^3)$ for obtaining the noise subspace projection matrix via \multicite{eq:EVD, eq:Pn, eq:MDL}, and ${O}(N_{\text{g}}K^2)$ for estimating the relative TO via \eqref{eq:TO_estmation_subspace}, where $N_{\text{g}}$ is the grid point number for searching. To ensure accurate alignment, $N_{\text{g}}\gg K$ is required.

\subsection{Covariance-Based Snapshot Alignment Procedure}
\label{sec_snapshot_alignment_covariance}

The subspace-based method described in Sec. \ref{sec_snapshot_alignment_subspace} does not assume correlation in the path gain across different snapshots. Consequently, it does not incorporate prior knowledge about the path gains, ${\boldsymbol{b}^{q}_t}$, derived from the aligned snapshots. This approach is suitable when the correlation across snapshots is negligible or unpredictable. 
On the other hand, in scenarios where the distribution of the path gain ${\boldsymbol{b}^{q}_t}$ is statistically stationary across snapshots, incorporating temporal correlation can be beneficial. Therefore, we also propose a covariance-based snapshot alignment method to effectively utilize this temporal correlation of path gains.

Since the PO $\varphi_{\!\text{o},t}^{q}$ often uniformly distributes on $[-\pi, \pi)$, the path gain shifted by PO, $\boldsymbol{b}_{t}^{q}\!\!\cdot\! \mathrm{e}^{\mathrm{j}\varphi_{\!\text{o},t}^{q}}$, is a zero-mean circular distributed complex random variable. We assume that $\boldsymbol{b}_{t}^{q}\!\!\cdot\! \mathrm{e}^{\mathrm{j}\varphi_{\!\text{o},t}^{q}}$ approximately follows a circular complex Gaussian distribution with a covariance matrix $\boldsymbol{R}_{\mathrm{b}}^{q}$. 
\begin{proposition}[Covariance-Based Relative TO Estimation]
	\label{prop:TO_estimation_covariance}
	With the approximation of stationary circular complex Gaussian distributed path gains $\boldsymbol{b}_{t}^{q}\!\cdot\! \mathrm{e}^{\mathrm{j}\varphi_{\!\text{o},t}^{q}}$, the estimate of relative TO $\Delta \tau_{\text{o},t}^{q}$ is given by 
	\vspace{-6pt}
	\begin{equation}
		\setlength{\jot}{-8pt}
		\begin{aligned}
			\!\!\!\widehat{\Delta \tau_{\text{o},t}^{q}} =& \arg\!\min_{\widetilde{\Delta\tau_{\text{o},t}^{q}}}
			\text{tr}\bigg(\!{\hat{\boldsymbol{h}}_t^q\,\!}^{\dagger} \!\cdot \text{diag}\Big(\!\boldsymbol{a}\big(\!\widetilde{\Delta\tau_{\text{o},t}^{q}}\!\big)\!\Big) \!\cdot\! \Big(\!\boldsymbol{R}_{\mathrm{w}}+\widehat{\sigma^2}\boldsymbol{I}\!\Big)^{\!\!-\!1} \\
			&\quad\quad\quad\quad\quad\quad\quad\quad\quad\quad \cdot \text{diag}\Big(\boldsymbol{a}^{\!*}\!\big(\widetilde{\Delta\tau_{\text{o},t}^{q}}\big)\!\Big) \!\cdot\! {\hat{\boldsymbol{h}}_t^q} \!\bigg) ,
			\label{eq:TO_estmation_covariance}
		\end{aligned}
		\vspace{-8pt}
	\end{equation}
	where $\boldsymbol{R}_{\mathrm{w}}$ is the shifted covariance matrix from \eqref{eq:cross-frequency-correlation}, and the noise power $\widehat{\sigma^2}$ can be determined empirically, or estimated as the mean of the eigenvalues corresponding to the noise subspace from \eqref{eq:EVD}.
	\newline\noindent\textit{Derivation:} see Appendix \ref{Apd:TO_estimation_covariance}. 
\end{proposition}

With the relative TO estimated as $\widehat{\Delta \tau_{\text{o},t}^{q}}$ using Proposition \ref{prop:TO_estimation_covariance}, the snapshot is then aligned similar to \eqref{eq:snapshot_alignment}. 

The complexity of a covariance-based standard snapshot alignment procedure includes ${O}(K^2)+{O}(K^3)$ for updating the covariance matrix and obtaining the matrix inverse, and ${O}(N_{\text{g}}K^2)$ for estimating the relative TO via \eqref{eq:TO_estmation_covariance}.

\subsection{Initial Snapshots Alignment Procedure}
\label{sec_initial_TO_alignment}
In this part of the discussion, we omit the CPI index superscript $q$ since the initial alignment procedure solely involves the first CPI.
The initial alignment procedure is specifically applied to the first $T_{\text{w}}$ snapshots to obtain $\boldsymbol{H}_{1:T_{\!\text{w}}}^{\text{A}}$. 
Direct application of the standard snapshot alignment procedure in Sec. \ref{sec_snapshot_alignment_subspace} and \ref{sec_snapshot_alignment_covariance} is impractical for these early snapshots, because the number of previously aligned snapshots is insufficient at this stage, leading to a singular covariance matrix. 

Since this algorithm aims to align the relative TO between snapshots, we bypass the alignment of the first snapshot by directly designating it as the aligned first snapshot, $\boldsymbol{H}_{1}^{\text{A}} \leftarrow \boldsymbol{h}_{1}$. This approach uses the TO of the first snapshot as the alignment reference, setting the initial TO residual as $\tau_{\text{o}}^{\text{r}} = \tau_{\text{o},1}$. 
The snapshots indexed from $K\!+\!1$ to $T_{\text{w}}$ are processed using the standard snapshot alignment procedures. By this stage, the number of aligned snapshots is at least $K$, ensuring a non-singular covariance matrix.

Next, we focus on aligning the snapshots indexed by $t\!\in\![2,\ldots, K]$. At this stage, the available aligned snapshots are given by $\big[\!\boldsymbol{H}^{\text{A}}\!\big]_{\!1:t-\!1} \!\!\in\!\! \mathbb{C}^{K\!\times\! (t\!-\!1)}$. Following the spatial smoothing technique \cite{spatial_smoothing, single_snapshot_MUSIC}, we divide the entire subcarriers into $K-s+1$ overlapping submatrices, each comprising $s$ subcarriers. By stacking the submatrix snapshots, we can estimate the submatrix signal subspace/covariance matrix. The detailed process is described below.

We construct the Hankel matrix of the aligned snapshots, $\boldsymbol{H}_{t-1}^{\text{HA}\!} \in \mathbb{C}^{s\!\times\!(K\!-\!s\!+\!1)(t\!-\!1)}$, and that of the $t$-th snapshot to be aligned, $\boldsymbol{h}_{t}^{{\text{H}}} \!\in\! \mathbb{C}^{s\!\times\!(K\!-\!s\!+\!1)}$,
\vspace{-3pt}
\begin{equation}
	\begin{aligned}
	\!\!\!\!\boldsymbol{H}_{t-1}^{\text{HA}} \!&= 
	\begin{bmatrix}
		\!\big[\!\boldsymbol{H}^{\text{A}\!}\big]_{\substack{\!\!\!\!1:s \\ \!1:t{\scalebox{0.7}{-\!}}1}} 
		& \!\!\big[\!\boldsymbol{H}^{\text{A}\!}\big]_{\substack{\!2:s{\scalebox{0.7}{+\!}}1 \\ \!\!\!1:t{\scalebox{0.7}{-\!}}1}} 
		& \!\!\!\!\!\cdots\!\!\! 
		& \!\!\big[\!\boldsymbol{H}^{\text{A}\!}\big]_{ \substack{\!{\scalebox{0.7}{$K$}}{\scalebox{0.7}{\!-}}s{\scalebox{0.7}{+\!}}1:{\scalebox{0.7}{$K$}}\! \\ \!\!\!\!\!\!\!\!\!1:t{\scalebox{0.7}{-\!}}1}}
	\end{bmatrix}\!\!\\
	\!\!\!\!\boldsymbol{h}_{t}^{\scalebox{0.6}{\text{H}\!}} &= \Big[
	\begin{matrix}
		\!\big[\!\boldsymbol{h}_{t}\big]_{\!1:s}
		& \!\!\big[\!\boldsymbol{h}_{t}\big]_{\!2:s{\scalebox{0.7}{+\!}}1} 
		& \!\!\cdots 
		& \!\!\big[\!\boldsymbol{h}_{t}\big]_ {\!{\scalebox{0.7}{$K$}}{\scalebox{0.7}{\!-}}s{\scalebox{0.7}{+\!}}1:{\scalebox{0.7}{$K$}}\!}
	\end{matrix}\Big],
	\end{aligned}
	\label{eq:Hankel}
	\vspace{-3pt}
\end{equation}
where $s=\lfloor \frac{(K+1)(t-1)}{t} \rfloor$ is chosen to retain the maximum equivalent bandwidth while ensuring that the submatrix correlation matrix is non-singular.
Each column of $\boldsymbol{h}_{t}^{{\text{H}}}$ is a submatrix snapshot, where the $k$-th is given by
\vspace{-3pt}
\begin{equation}
	{\big[\boldsymbol{h}_{t}^{{\text{H}}}\big]_{\!k}} \!\!=\! \text{diag}\!\Big(\!\big[\boldsymbol{a}\!(\Delta \tau_{\text{o},t})\big]_{\!1:s}\!\Big) \!{\boldsymbol{A}^{\text{H}\text{r}}} \!\cdot \boldsymbol{D}_{\!k} \cdot \boldsymbol{b}_t' \!\cdot  \mathrm{e}^{\mathrm{j}(\!\varphi_{\text{o},t}-{\scalebox{0.6}{2}}\pi k\Delta\!f\tau_{\!\text{o},\!t}\!)} \!+\! \boldsymbol{z},\!
	\vspace{-3pt}
\end{equation}
where ${\boldsymbol{A}^{\!\text{H}\text{r}}}\!\!\!=
\!\!\text{diag}\!\Big(\! \big[\!\boldsymbol{a}\!(\!\tau_{\!\text{o}}^{\text{r}})\!\big]_{{\scalebox{0.6}{\!\!1\!:}}s}\!\Big) \!
\Big[\!
\big[\!\boldsymbol{a}\!(\!\tau_{\!\text{s},\!1}\!)\!\big]_{{\scalebox{0.6}{\!\!1\!:}}s} {\scalebox{0.8}{,..,\!}}
\big[\!\boldsymbol{a}\!(\!\tau_{\!\text{s},\!L_{\text{s}}}\!)\!\big]_{{\scalebox{0.6}{\!\!1\!:}}s},\!
\big[\!\boldsymbol{a}\!(\!\tau_{\!1}\!)\!\big]_{{\scalebox{0.6}{\!\!1\!:}}s} {\scalebox{0.8}{,..,\!}}
\big[\!\boldsymbol{a}\!(\!\tau_{\!L_{\text{d}}}\!)\!\big]_{{\scalebox{0.6}{\!\!1\!:}}s}\!
\Big]$, 
$\boldsymbol{D}_{\!k}\!\!=\!\!\text{diag}\!\Big({\scalebox{0.7}{\!\!}}\big[
\mathrm{e}^{{\scalebox{0.7}{-}}\mathrm{j}{\scalebox{0.6}{2}}\pi k\Delta\!f\tau_{\!\text{s},\!1}} \!{\scalebox{0.8}{\!,..,}} 
\mathrm{e}^{{\scalebox{0.7}{-}}\mathrm{j}2\pi k\Delta\!f\tau_{\!\text{s},\!L_{\text{s}}}}\!\!,\!
\mathrm{e}^{{\scalebox{0.7}{-}}\mathrm{j}2\pi k\Delta\!f\tau_{\!1}} \!{\scalebox{0.8}{\!,..,}}
\mathrm{e}^{{\scalebox{0.7}{-}}\mathrm{j}2\pi k\Delta\!f\tau_{\!L_{\text{d}}}}\!
\big]{\scalebox{0.7}{\!\!}}\Big)$, and 
$\boldsymbol{b}_t'\!\!=\!\!\big[\!\beta_{\text{s},\!1\!},\!..,\beta_{\text{s},\!L_{\text{s}}\!},\beta_{\text{d},\!1\!},\!..,\beta_{\text{d},\!L_{\text{d}}}\!\big]^{\!\intercal\!\!}$.
With $\boldsymbol{D}_{\!k} \!\cdot\! {\boldsymbol{b}_t} \!\cdot\! \mathrm{e}^{\mathrm{j}(\!\varphi_{\text{o},\!t}{{\scalebox{0.7}-}}{\scalebox{0.6}{2}}\pi k\Delta\!f\tau_{\!\text{o},\!t}\!)}\!$ treated as the path gain varying across submatrix snapshots, the submatrix signal subspace is the column span of ${\boldsymbol{A}^{\text{H}\text{r}}}$. 
Therefore, for the initial step of subspace-based method, the nullspace projection matrix of the aligned snapshots for submatrices, ${\boldsymbol{P}_{\mathrm{n}}^{\text{H}\text{r}}}$, can be estimated via \multicite{eq:covariance, eq:EVD, eq:Pn, eq:MDL}, with $\boldsymbol{H}^{\text{w}}$ replaced by $\boldsymbol{H}_{t-1}^{\text{HA}}$. 
The log-likelihood of the snapshot to be aligned for the subspace-based method is given by the sum of that for each submatrix, 
\vspace{-5pt}
\begin{equation}
	\!\!\ln\!\!\bigg(\!p\Big(\boldsymbol{h}_{t}^{{\text{H}}} ; \Delta\tau_{\text{o},t},{\boldsymbol{A}^{\text{H}\text{r}}}\!\Big)\!\!\bigg)\!\!
	=\!\!\! \sum_{k=1}^{K\!-\!s\!+\!1}\!\!\ln\!\!\Bigg(\!p\bigg(\!{\Big[\boldsymbol{h}_{t}^{{\text{H}}}\Big]_{\!k}} ; \Delta\tau_{\text{o},t},{\boldsymbol{A}^{\text{H}\text{r}}}\!\bigg)\!\!\Bigg)\!.
	\vspace{-3pt}
\end{equation}
Thus, the relative TO $\Delta\tau_{\text{o},t}$ can be estimated via \eqref{eq:TO_estmation_subspace}, with $\widehat{\boldsymbol{P}_{\mathrm{n}}^{\,\text{r},q}}$ and ${\boldsymbol{h}_t^q}$ replaced by $\widehat{{\boldsymbol{P}_{\mathrm{n}}^{\,\text{H}\text{r}}}}$ and $\boldsymbol{h}_{t}^{{\text{H}}}$. 

The case for the covariance-based method is similar to that of the subspace-based one. The covariance matrix of the submatrix snapshots is given by ${\boldsymbol{R}_{\text{h}}^{\,\text{H}\text{r}}}\!\!=\!{\boldsymbol{A}^{\text{H}\text{r}}} \Big(\!\sum_{k=1}^{K{\scalebox{0.8}{\!-}}s{\scalebox{0.7}{+}}1}\! \boldsymbol{D}_k \boldsymbol{R}_{\mathrm{b}} \boldsymbol{D}_k^{\dagger}\Big) {\boldsymbol{A}^{\text{H}\text{r}}}\!+\sigma^2\boldsymbol{I}$, which can be estimated as \eqref{eq:covariance}, with $\boldsymbol{H}^{\text{w}}$ replaced by $\boldsymbol{H}_{t-1}^{\text{HA}}$; and $\Delta\tau_{\text{o},t}$ can be estimated via \eqref{eq:TO_estmation_covariance}, with $\boldsymbol{R}_{\mathrm{w}}$ and ${\boldsymbol{h}_t^q}$ replaced by $\boldsymbol{H}_{t-1}^{\text{HA}} {\boldsymbol{H}_{t-1}^{\text{HA}}}^{\dagger}$ and $\boldsymbol{h}_{t}^{{\text{H}}}$. 

\renewcommand\algorithmicindent{1em}
\begin{algorithm}[t]
	\caption{Algorithm for TO Alignment within CPI}
	\label{alg:TO_alignment}
	\begin{algorithmic}
		\Require Channel snapshots with snapshot-specific TO, $\hat{\boldsymbol{h}}_{t}$ for $t \in [1,\ldots,\infty]$. 
		\Ensure Aligned channel snapshots, $\boldsymbol{h}^{\text{A}}_t$. For two aligned snapshots within the same CPI, the remaining TO residuals are approximately the same. 
		\Procedure{Main}{}
		\State Initialize $\boldsymbol{H}^{\text{w}}\!\gets\!\boldsymbol{0}_{K\!\times\! T_{\text{w}}}$ \grayComment{to store the sliding window}
		\State Assign $\boldsymbol{h}^{\text{A}}_1 \!\!\leftarrow \hat{\boldsymbol{h}}_{1}$, $\boldsymbol{H}^{\text{w}}\!(:,1) \leftarrow \hat{\boldsymbol{h}}_{1}$ \grayComment{skip $1^{\text{st}}$ snapshot}
		\State Initialize $t \gets 2$ \grayComment{Index of the snapshot to be aligned}
		
		\While{comes a channel snapshot measurement $\boldsymbol{h}_t$}
		\If{$t\leq K$} \grayComment{Initial snapshot alignment}
		\State Construct $\boldsymbol{H}_{t-1}^{{\text{HA}}}$ and $\boldsymbol{h}_{t}^{\scalebox{0.6}{\text{H}\!}}$ according to \eqref{eq:Hankel}
		\EndIf
		\State \textbf{Call:} 
		\Statex \hspace{3em}\textbf{procedure} \Call{1.A}{} for subspace-based method; or
		\Statex \hspace{3em}\textbf{procedure} \Call{1.B}{} for covariance-based method
		\State $\boldsymbol{H}^{\text{w}} \!\gets\! \Big[\boldsymbol{H}^{\text{w}}\!(:,2\!\!:\!T_{\text{w}}), \boldsymbol{h}^{\text{A}}_t\Big]$ \grayComment{Update $\boldsymbol{H}^{\text{w}}$}
		\State \textbf{Yield} $\boldsymbol{h}^{\text{A}}_t$ \grayComment{return the aligned snapshot and proceed}
		\State $t\leftarrow t+1$ \grayComment{Update counter}
		\EndWhile
		\EndProcedure
		\vspace{-6pt}
		
		\Statex
		\Procedure{1.A}{} \grayComment{Subspace-based snapshot alignment}
		\State Apply eigenvalue decomposition with eigenvalues in descend order, 			$\boldsymbol{H}^{\text{w}}{\boldsymbol{H}^{\text{w}}}^{\dagger}=\boldsymbol{U}\boldsymbol{\Lambda}\boldsymbol{U}^{\dagger}$
		\State Estimate signal subspace dimension $\hat{d}$ with $\boldsymbol{\Lambda}$ via \eqref{eq:MDL}
		\State Construct null-space projection $\widehat{\boldsymbol{P}_{\!\mathrm{n}}^{\text{r}}} \!\!\gets\!\! \boldsymbol{U}\!(:,\hat{d}\!+\!\!1\!) {\boldsymbol{U}\!(:,\hat{d}\!\!+\!\!1\!)}^{\!\dagger}$
		\State Estimate relative TO $\widehat{\Delta\tau_{\text{o},t}}$ with $\widehat{\boldsymbol{P}_{\mathrm{n}}^{\,\text{r}}}$ and ${\hat{\boldsymbol{h}}_t}$ via \eqref{eq:TO_estmation_subspace}
		\State $\boldsymbol{h}_{t}^{\text{A}} \gets \text{diag}\big(\boldsymbol{a}^{\!*}\!(\widehat{\Delta\tau_{\text{o},t}})\!\big) \!\cdot\! {\hat{\boldsymbol{h}}_t}$
		\EndProcedure
		\vspace{-6pt}
		
		\Statex
		\hspace{-1em}\Procedure{1.B}{} \grayComment{Covariance-based snapshot alignment}
		\State ${\boldsymbol{R}_{\text{w}}} \gets \boldsymbol{H}^{\text{w}} {\boldsymbol{H}^{\text{w}}}^{\dagger}$
		\State Estimate relative TO $\widehat{\Delta\tau_{\text{o},t}}$ with ${\boldsymbol{R}_{\text{w}}}$ and ${\hat{\boldsymbol{h}}_t}$ via \eqref{eq:TO_estmation_covariance}
		\State $\boldsymbol{h}_{t}^{\text{A}} \gets \text{diag}\big(\boldsymbol{a}^{\!*}\!(\widehat{\Delta\tau_{\text{o},t}})\!\big) \!\cdot\! {\hat{\boldsymbol{h}}_t}$
		\EndProcedure
	\end{algorithmic}
\end{algorithm}

Due to the reduction in the equivalent bandwidth of the submatrix compared to the entire subcarriers and the increase in the number of equivalent paths (as the static paths can no longer be considered as a single merged path), the TO alignment errors in the initial procedures are expected to be larger than those in standard procedures. However, as the standard snapshot alignment procedure iterates, the aligned snapshots obtained in the initialization step are replaced in the sliding window, and this effect gradually diminishes. 

\subsection{Algorithm Acceleration via SFFT}
\label{sec_TO_alignment_acceleration}

To facilitate real-time processing, we implement SFFT \cite{MUSIC_FFT, SFFT} to accelerate the TO alignment algorithm, particularly in the search process via \multicite{eq:TO_estmation_subspace, eq:TO_estmation_covariance}. 
Considering the search process that employs a total number of grid points $N_{\text{g}}$, it is typically necessary to set $N_{\text{g}}\gg K$ to ensure negligible discretization error. The proposed algorithm acceleration via SFFT can reduce the complexity of the search process from $O(N_{\text{g}}K^2)$ to $O(N_{\text{g}}K\log(K))$, leading to an overall algorithmic complexity of $O(K^3)+O(N_{\text{g}}K\log(K))$ for processing a channel snapshot.

For the subspace-based method, SFFT acceleration is implemented by 
\vspace{-6pt}
\begin{equation}
	\boldsymbol{S}=\boldsymbol{\mathrm{W}} \cdot \text{diag}(\boldsymbol{h}_{t}^{*}) \cdot {\widehat{\boldsymbol{P}_{\text{n}}^{\text{r},q}}}^{\frac{1}{2}},
	\vspace{-3pt}
\end{equation}
where $\boldsymbol{\mathrm{W}} \!\in\! \mathbb{C}^{N_{\text{g}} \!\times\! K}$ is a weighting matrix that incorporates zero-padding and DFT, and its multiplication is computed using SFFT. For the covariance-based method, ${\widehat{\boldsymbol{P}_{\text{n}}^{\text{r},q}}}^{\frac{1}{2}}$ is replaced by $\big({\boldsymbol{R}_{\mathrm{w}}}+\widehat{\sigma^2}\boldsymbol{I}\big)^{\!\!-\frac{1}{2}}$, both constructed directly through eigenvalue decomposition with a complexity of $O(K^3)$. The relative TO $\Delta\tau_{\text{o},t}^{q}$ is then estimated by identifying the minimum of the spectrum $S(n)=\Big| \big[\boldsymbol{S}^{\intercal}\big]_{\!n} \Big|^2$. 

The complete relative TO alignment algorithm is summarized in Algorithm \ref{alg:TO_alignment}.

\section{TO Residual Compensation \& Delay Estimation}
\label{sec:TO_compensation_delay_estimation}

The relative TO alignment algorithm discussed above enables the application of signal processing techniques based on snapshot stacking, such as subspace methods. 
However, at this stage, delay estimation still faces two problems: 1) the aligned channel snapshots, $\boldsymbol{H}^{\text{A},\,q}$, still contain the TO residual $\tau_{\text{o}}^{\text{r},q}$, which hinders the estimation of paths' absolute delay; and 2) the signal contains the static paths component term $\boldsymbol{h}_{\text{s}}\cdot\boldsymbol{1}_{T\!\times\! 1}^{\intercal}$, which can interfere with the identification of dynamic
paths. 

To address these problems, this section first proposes a \textit{TO Residual Compensation} scheme by integrating a \textit{Reference Static Channel Component}. Note that this reference is necessary for absolute delay estimation, since channel snapshots $\hat{\boldsymbol{H}}^{q}$ measured solely by the BS are inadequate for this task \cite{Performance_Bound_Asyn}. The proposed scheme involves only a limited number of dedicated channel measurements, minimizing the impact on communication performance, and is effective in NLoS environments. Subsequently, we introduce a \textit{Static-Clutter-Suppressed Delay Estimation}, which further utilizes the reference static channel component to suppress the interference from the static paths components, and robustly estimate the delays of dynamic paths. 

\begin{figure}[t]
	\vspace{-12pt}
	\includegraphics[width=9cm, height=5.3cm]{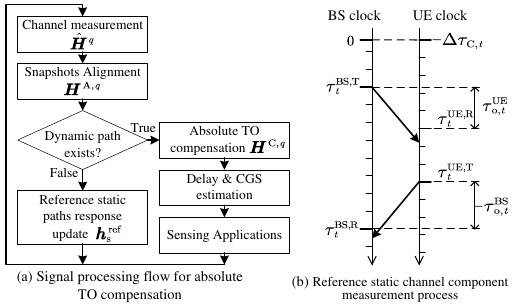}
	\vspace{-18pt}
	\caption{Diagram of the absolute TO compensation based on reference static paths response. (a) Signal processing flow for absolute TO compensation; (b) Reference static paths response measurement process.}
	\label{Fig_TO_comp}
	\vspace{-3pt}
\end{figure}

Specifically, we first implement an auto-calibration step based on bidirectional channel measurements, which is regularly executed when the environment is free of dynamic targets, to obtain and update a reference static paths response, $\!\boldsymbol{h}_{\text{s}}^{\!\text{ref}}\!$, as shown in Fig. \ref{Fig_TO_comp}.a. 
This reference $\boldsymbol{h}_{\text{s}}^{\!\text{ref}}$ is a complex-scaled version of the static paths component,  $\boldsymbol{h}_{\text{s}}^{\!\text{ref}}\!\!=\!\!\beta_{\text{s}}^{\text{ref}}\!\cdot\!\boldsymbol{h}_{\text{s}},\, \beta_{\text{s}}^{\text{ref}}\!\!\in\!\!\mathbb{C}$, containing absolute delay information of the merged static path. 
\mbox{Then, we design a TO residual} estimation and compensation procedure for use during the detection of dynamic targets. 
This procedure estimates and compensates for the TO residual $\tau_{\!\text{o}}^{\!\text{r},q}$ from the aligned snapshots $\!\boldsymbol{H}^{\text{A},q}\!\!$, thereby obtaining the compensated snapshots, $\!\boldsymbol{H}^{\text{C},q}\!\!$, which include the absolute delay information of the paths. 
Subsequently, the estimates of the dynamic paths, $\widehat{\tau}^{q}_{1},\!..,\widehat{\tau}^{q}_{L_{\text{d}}}$, are obtained in the delay estimation procedure, where the reference $\boldsymbol{h}_{\text{s}}^{\text{ref}}$ is further used for static clutter suppression. 

Note that without the bi-directional channel measurement step (which may require a dedicated protocol), the framework can still obtain an alternative reference static paths response, which contains a fixed initial TO. Using this alternative reference, the subsequent delay and CGS estimation procedures (Sec. \ref{sec:TO_residual_compensation_B}, \ref{sec:delay_estimation}, and \ref{sec:CGS_estimation_extension}) can also proceed, with resulting delay estimates shifted by the initial TO. Despite this, the relative delay between paths, delay changes over time, and CGS (and thus Doppler) information can still be accurately estimated.

In the following, we detail the process of the reference static paths response acquisition and TO residual compensation for aligned CPIs. 

\subsection{Acquisition of Reference Static Paths Response}
\label{sec:TO_residual_Compensation_A}
The reference static paths response is acquired and updated through multiple repetitions of the bidirectional channel measurements shown in Fig. \ref{Fig_TO_comp}b, conducted during a period without dynamic paths. 

Consider $T_{\text{s}}$ bidirectional channel measurements are conducted. As shown in Fig. \ref{Fig_TO_comp}b, for the $t$-th measurement, $\Delta\tau_{\text{C},t}$ represents the clock error where the UE lags behind the BS. The BS transmits a packet at the local clock $\tau_{t}^{\text{BS},\text{T}}$, and the UE detects the packet boundary and begins measuring the channel snapshot $\boldsymbol{h}^{\text{UE}}_{t}$ at its local clock $\tau_{t}^{\text{UE},\text{R}}$. Subsequently, the UE sends a packet, and the BS detects the packet and starts measuring the channel snapshot $\boldsymbol{h}^{\text{BS}}_{t}$ at $\tau_{t}^{\text{UE},\text{T}}$ and $\tau_{t}^{\text{BS},\text{R}}$. Due to the short interval between one complete round of sending and receiving, the variation in the clock error $\Delta\tau_{\text{C},t}$ between channel snapshots at the BS and UE (caused only by clock drift) is negligible. Therefore, in the $t$-th bidirectional channel measurement, the TO in the channel snapshots at the BS and UE is given by
\vspace{-9pt}
\begin{equation}
	\quad\quad\left\{
	\begin{array}{c}
		\!\!\tau_{\text{o},t}^{\text{BS}}=\tau_{t}^{\text{BS},\text{R}}-\tau_{t}^{\text{UE},\text{T}}-\Delta\tau_{\text{C},t}\\
		\!\tau_{\text{o},t}^{\text{UE}}=\tau_{t}^{\text{UE},\text{R}}-\tau_{t}^{\text{BS},\text{T}}+\Delta\tau_{\text{C},t} \,.
	\end{array}
	\right.
	\vspace{-3pt}
\end{equation}

Apply snapshot alignment (Algorithm \ref{alg:TO_alignment}) to the channel snapshots at BS, $\boldsymbol{H}^{\text{BS}}=\big[\boldsymbol{h}^{\text{BS}}_{1},\ldots,\boldsymbol{h}^{\text{BS}}_{T_{\text{s}}}\big]$, this yields relative TO estimates $\Delta\tau_{\text{o},t}^{\text{BS}}, t\in[1,\ldots,T_{\text{s}}]$ and the aligned snapshots $\boldsymbol{H}^{\text{A},\text{BS}}$. Then, apply EVD to the covariance matrix of the aligned snapshots, $\boldsymbol{H}^{\text{A},\text{BS}} {\boldsymbol{H}^{\text{A},\text{BS}}}^{\dagger}=\boldsymbol{U}^{\text{BS}}\boldsymbol{\Sigma}^{\text{BS}}{\boldsymbol{U}^{\text{BS}}}^{\dagger}$, and the static paths component in the channel measurement at BS, shifted by the TO residual $\tau_{\text{o}}^{\text{r},\text{BS}}$, can be obtained as the eigenvector corresponding to the largest eigenvalue, $\boldsymbol{h}_{\text{s}}^{\text{r}, \text{BS}}=\big[\boldsymbol{U}^{\text{BS}}\big]_{1}$. 
Similarly, from the channel measurement at UE, relative TO estimates $\Delta\tau_{\text{o},t}^{\text{UE}}, t\in[1,\ldots,T_{\text{s}}]$ and the static paths component $\boldsymbol{h}_{\text{s}}^{\text{r}, \text{UE}}$, which is shifted by TO residual $\tau_{\text{o}}^{\text{r},\text{UE}}$, are obtained. 

Denote the average of $\Delta\tau_{\text{C},t}$ in the total $T_{\text{s}}$ measurements as $\Delta\tau_{\text{C}}$, $\Delta\tau_{\text{C}}\!\!=\!\!\frac{1}{T_{\text{s}}}\!\! \sum_{t=1}^{T_{\text{s}}} \Delta\tau_{\text{C},t}$.
Given $\Delta\tau_{\text{C}}$, the TO residual $\tau_{\text{o}}^{\text{r},\text{BS}}$ in $\boldsymbol{h}_{\text{s}}^{\text{r}, \text{BS}}$ and the TO residual $\tau_{\text{o}}^{\text{r},\text{UE}}$ in  $\boldsymbol{h}_{\text{s}}^{\text{r}, \text{UE}}$ are estimated as
\vspace{-5pt}
\begin{equation}
	\!\!\!\!\left\{\!\!
	\begin{array}{c}
		\!\!\!\widehat{\tau_{\text{o}}^{\text{r},\text{BS}}}\big({\Delta\tau_{\text{C}}}\big)\!=-{\Delta\tau_{\text{C}}}\!+\!\frac{1}{T_{\text{s}}}\sum\limits_{t=1}^{T_{\text{s}}}\big(\tau_{t}^{\text{BS},\text{R}}\!-\!\tau_{t}^{\text{UE},\text{T}}\!-\!\Delta\tau_{\text{o},t}^{\text{BS}}\big)\\
		\!\!\widehat{\tau_{\text{o}}^{\text{r},\text{UE}}}\big({\Delta\tau_{\text{C}}}\big)\!=+{\Delta\tau_{\text{C}}} \!+\!\frac{1}{T_{\text{s}}}\sum\limits_{t=1}^{T_{\text{s}}} \big(\tau_{t}^{\text{UE},\text{R}}\!-\!\tau_{t}^{\text{BS},\text{T}}\!-\!\Delta\tau_{\text{o},t}^{\text{UE}}\big).
	\end{array}\!\!
	\right.
	\vspace{-3pt}
\end{equation}
Utilizing channel reciprocity, the unknown clock error parameter ${\Delta\tau_{\text{C}}}$ can be estimated by maximizing the similarity between the channels at the BS and UE, 
\vspace{-3pt}
\begin{equation}
	\setlength{\jot}{-2pt}
	\begin{aligned}
	\widehat{{\Delta\tau_{\text{C}}}} =& \arg\!\max_{\widetilde{{\Delta\tau_{\text{C}}}}} 
	\bigg|{\boldsymbol{h}_{\text{s}}^{\text{r}, \text{UE}}}^{\dagger} \!\!\cdot\!
	\text{diag}\!\bigg(\!\boldsymbol{a}\!\Big(\!\widehat{\tau_{\text{o}}^{\text{r},\text{UE}}}\!\big(\widetilde{{\Delta\tau_{\text{C}}}}\!\big)\!\Big)\!\bigg) \\ & \quad\quad\quad\quad\quad\quad\quad\cdot\!
	\text{diag}\!\bigg(\!\boldsymbol{a}^{\!*}\!\Big(\!\widehat{\tau_{\text{o}}^{\text{r},\text{BS}}}\!\big(\widetilde{{\Delta\tau_{\text{C}}}}\!\big)\!\Big)\!\bigg) \!\!\cdot\!
	\boldsymbol{h}_{\text{s}}^{\text{r}, \text{BS}}\bigg|
	\end{aligned}
	\vspace{-3pt}
\end{equation}
and the reference static paths response is estimated as 
\vspace{-3pt}
\begin{equation}
	\boldsymbol{h}_{\text{s}}^{\text{ref}} \gets \text{diag}\!\bigg(\boldsymbol{a}^{\!*}\!\Big(\widehat{\tau_{\text{o}}^{\text{r},\text{BS}}}\!\big(\widehat{\Delta\tau_{\text{C}}}\!\big)\!\Big)\!\bigg) \cdot \boldsymbol{h}_{\text{s}}^{\text{r}, \text{BS}}. 
	\vspace{-3pt}
\end{equation}
\vspace{-12pt}

\subsection{TO Residual Compensation}
\label{sec:TO_residual_compensation_B}
We introduce a TO residual compensation algorithm based on the subspace principle, which utilizes the reference static paths response $\boldsymbol{h}_{\text{s}}^{\text{ref}}$ to estimate and compensate for the TO residual $\tau_{\text{o}}^{\text{r},q}$ from the aligned channel snapshots in the $q$-th CPI, $\boldsymbol{H}^{\text{A},\,q}$, obtaining the channel snapshots $\boldsymbol{H}^{\text{C},\,q}$ that contains the absolute path delay information. 

The foundation of the TO residual compensation algorithm is that $\boldsymbol{h}_{\text{s}}$ is always in the signal subspace and is thus orthogonal to the noise subspace. Leveraging the aligned channel snapshots $\boldsymbol{H}^{\text{A},\,q}$ obtained by Algorithm \ref{alg:TO_alignment}, we estimate the projection matrix, $\widehat{\boldsymbol{P}_{\mathrm{n}}^{\,\text{r},q}}$, which projects the signal into the noise subspace shifted by the TO residual, as adapted from \multicite{eq:covariance, eq:EVD, eq:Pn, eq:MDL}. Given the estimate of the TO residual, $\widehat{\tau_{\text{o}}^{\text{r},q}}$, we can derive from \eqref{eq:A_shifted} and \eqref{eq:projection_shifted} that the projection matrix of the non-shifted noise subspace can be expressed as 
\vspace{-6pt}
\begin{equation}
\widehat{\boldsymbol{P}_{\mathrm{n}}^{q}}\Big(\widehat{\tau_{\text{o}}^{\text{r},q}}\Big)= \text{diag}\bigg(\!\boldsymbol{a}\Big(\widehat{\tau_{\text{o}}^{\text{r},q}}\Big)\!\bigg) \!\cdot\! \widehat{\boldsymbol{P}_{\mathrm{n}}^{\,\text{r},q}} \cdot \text{diag}\bigg(\boldsymbol{a}^{\!*}\!\Big(\widehat{\tau_{\text{o}}^{\text{r},q}}\Big)\!\bigg).
\vspace{-6pt}
\end{equation}
Therefore, by exploiting the orthogonality between $\boldsymbol{h}_{\text{s}}$ and the non-shifted noise subspace, we estimate the TO residual ${\tau_{\text{o}}^{\text{r},q}}$ by minimizing the projection of $\boldsymbol{h}_{\text{s}}$ into the non-shifted noise subspace, leading to 
\vspace{-6pt}
\begin{equation}
\begin{aligned}
	\widehat{\tau_{\text{o}}^{\text{r},q}} =& \arg\!\min_{\widetilde{\tau_{\text{o}}^{\text{r},q}}}
	\text{tr}\bigg(\!{\boldsymbol{h}_{\text{s}}^{\text{ref}}\,\!}^{\dagger} \!\!\cdot\! \text{diag}\!\Big(\!\boldsymbol{a}\!\big(\widetilde{\tau_{\text{o}}^{\text{r},q}}\big)\!\Big) \!\!\cdot\! \widehat{\boldsymbol{P}_{\mathrm{n}}^{\,\text{r},q}} \!\!\cdot\! \text{diag}\!\Big(\!\boldsymbol{a}^{\!*}\!\big(\widetilde{\tau_{\text{o}}^{\text{r},q}}\big)\!\Big) \!\!\cdot\! {\boldsymbol{h}_{\text{s}}^{\text{ref}}} \!\bigg) \!,\!
\end{aligned}
\vspace{-3pt}
\label{eq:h_ref}
\end{equation}
and subsequently, the TO residual is compensated from the aligned snapshots as  
\vspace{-5pt}
\begin{equation}
	\boldsymbol{H}^{\text{C},\,q} \gets \text{diag}\Big(\boldsymbol{a}^{\!*}\!\big(\widehat{\tau_{\text{o}}^{\text{r},q}}\big)\!\Big) \!\cdot\!  \boldsymbol{H}^{\text{A},\,q}. 
	\vspace{-6pt}
\end{equation}

\subsection{Static-Clutter-Suppressed Delay Estimation}
\label{sec:delay_estimation}
Dynamic path delays are estimated using principles similar to the MUSIC algorithm \cite{MUSIC}. Delay estimation is unaffected by the PO, as the PO is canceled out during the computation of the covariance matrix. However, the static clutter, which is typical in indoor environments, poses a major challenge for delay estimation. 
Compared to the traditional delay estimation problem, the signal subspace includes an additional basis vector, the frequency response vector of the merged static path $\boldsymbol{h}_{\text{s}}$. Consequently, when a dominant path exists in the static paths component, a corresponding peak appears in the MUSIC spectrum, interfering with the identification of dynamic paths.
To address this issue, we propose a modified MUSIC spectrum that suppresses the interference from static paths by utilizing the reference static paths response $\boldsymbol{h}_{\text{s}}^{\text{ref}}$. 

Referring to the classic MUSIC algorithm \cite{MUSIC}, the orthonormal generator matrix of the signal subspace, $\boldsymbol{U}_{\text{S}}$, is first obtained via the method introduced in Sec. \ref{sec:subspace_preliminary}, using the TO-compensated CSI matrix $\boldsymbol{H}^{\text{C},q}$.  The projection matrix into the noise subspace, $\boldsymbol{P}_{\text{n}}$, is obtained as $\boldsymbol{P}_{\text{n}}=\boldsymbol{I}-\boldsymbol{U}_{\text{S}}\boldsymbol{U}_{\text{S}}^{\dagger}$. 
The  modified MUSIC spectrum is then obtained as
\vspace{-4pt}
\begin{equation}
	S_{\tau}(\tilde{\tau}) = \frac{\boldsymbol{a}(\tilde{\tau})^{\dagger}\cdot\boldsymbol{P}_{\text{n},\boldsymbol{h}_{\text{s}}}\cdot\boldsymbol{a}(\tilde{\tau})}{\boldsymbol{a}(\tilde{\tau})^{\dagger}\cdot\boldsymbol{P}_{\text{n}}\cdot\boldsymbol{a}(\tilde{\tau})},
	\label{eq:modified_MUSIC}
	\vspace{-4pt}
\end{equation}
where $\boldsymbol{P}_{\text{n},\boldsymbol{h}_{\text{s}}}$ is the projection matrix into the orthogonal space of the merged static path, given by $\boldsymbol{P}_{\text{n},\boldsymbol{h}_{\text{s}}}=\boldsymbol{I}-\nicefrac{\boldsymbol{h}_{\text{s}}^{\text{ref}}{\boldsymbol{h}_{\text{s}}^{\text{ref}}}^{\dagger}}{\big|\boldsymbol{h}_{\text{s}}^{\text{ref}}\big|^2}$.
The dynamic paths delay estimates, $\widehat{\tau}^{q}_{1},\ldots,\widehat{\tau}^{q}_{L_{\text{d}}}$, are given as the $L_{\text{d}}$ largest peaks of the modified MUSIC spectrum $S_{\tau}(\tilde{\tau})$. 
Intuitively, by computing the ratio of two spectrums, where $\nicefrac{1}{\boldsymbol{a}(\tilde{\tau})^{\!\dagger}\!\cdot\!\boldsymbol{P}_{\!\text{n}\!,\boldsymbol{h}_{\text{s}}}\!\cdot\boldsymbol{a}(\tilde{\tau})}$ reflects the peaks caused by the merged static path and $\nicefrac{1}{\boldsymbol{a}(\tilde{\tau})^{\!\dagger}\!\cdot\!\boldsymbol{P}_{\!\text{n}}\!\cdot\boldsymbol{a}(\tilde{\tau})}$ reflects peaks due to both the merged static path and dynamic target paths, the modified MUSIC spectrum effectively suppresses the interference from static clutter and preserves the peaks of dynamic target paths.

\section{CGS Estimation \& Extension to MIMO-ISAC}
\label{sec:CGS_estimation_extension}
The CSI matrix after TO compensation can be expressed as 
\vspace{-7pt}
\begin{equation}
\boldsymbol{H}^{\text{C},q}=\big[\boldsymbol{h}_{\text{s}}\, \boldsymbol{A}^{q}_{\text{d}}\big] \cdot \big[\boldsymbol{1}_{T\!\times\! 1}\, \boldsymbol{B}^{q}_{\text{d}}\big]^{\intercal}\cdot\text{diag}\big(\boldsymbol{\varphi}_{\text{o}}^{q}\big)+\boldsymbol{z},
\vspace{-3pt}
\end{equation}
which still contains the snapshot-specific PO $\boldsymbol{\varphi}_{\text{o}}^{q}$, which obscures the phase information of the CGS. In this section, we introduce the \textit{PO Compensation and CGS Estimation} algorithm to estimate the CGS of each dynamic path in the existence of PO. Additionally, we extend the framework from single-antenna systems to MIMO-ISAC.

\subsection{PO Compensation and CGS Estimation}
\label{sec:CGS_estimation}

The framework utilizes the TO-compensated CSI snapshots, $\boldsymbol{H}^{\text{C},q}$, and the dynamic path delay estimates, $\widehat{\tau}^{q}_{1},\ldots,\widehat{\tau}^{q}_{L_{\text{d}}}$, to separate and recover the CGS, $\boldsymbol{\beta}^{q}_{l}$, for each target. The basic idea is to leverage the achieved high delay resolution to separate the signals from different paths, and utilize the static paths component to estimate and compensate for the PO.

Specifically, the signal is separated by 
\vspace{-3pt}
\begin{equation}
	\boldsymbol{H}^{\text{sep},q}=\begin{bmatrix}
		{\widehat{\boldsymbol{A}_{\text{d}}^{q}}}^{+}\\
		{\widehat{\boldsymbol{A}_{\text{d,null}}^{q}}}^{\intercal}
	\end{bmatrix}
	\cdot\boldsymbol{H}^{\text{C},q},
	\label{eq:signal_sep}
	\vspace{-8pt}
\end{equation}
where $\widehat{\boldsymbol{A}_{\text{d}}^{q}}\!=\!\big[\boldsymbol{a}\!(\hat{\tau}^{q}_1),\!..,\boldsymbol{a}\!(\hat{\tau}^{q}_{L_{\text{d}}})\big]\!\in\!\mathbb{C}^{\!K\!\times\!L_{\text{d}}}$ is the estimate of the dynamic paths frequency response matrix; ${\widehat{\boldsymbol{A}_{\text{d}}^{q}}}^{+}$ is the pseudo-inverse of $\widehat{\boldsymbol{A}_{\text{d}}^{q}}$; and $\widehat{\boldsymbol{A}_{\text{d,null}}^{q}}$ is the orthonormal kernel matrix of $\widehat{\boldsymbol{A}_{\text{d}}^{q}}^{\intercal}$. The result is structured as
\vspace{-3pt}
\begin{equation}
	\boldsymbol{H}^{\text{sep},q} = \big[\boldsymbol{\beta}_{1}^{\text{PO},q},\ldots,\boldsymbol{\beta}_{L_{\text{d}}}^{\text{PO},q},{\boldsymbol{H}^{{\text{proj},q}}}^{\intercal}\big]^{\intercal},
	\vspace{-6pt}
\end{equation}
where the first $L_{\text{d}}$ rows are the separated signal for the dynamic paths. For the $l$-th dynamic path, ${\boldsymbol{\beta}_{l}^{\text{PO},q}} = \text{diag}(\text{e}^{\text{j}\boldsymbol{\varphi}_\text{o}^{q}})\cdot\big(\boldsymbol{\beta}_{l}^q+\beta_l^{\text{r},q}\cdot\boldsymbol{1}_{T\times 1}\big)+\boldsymbol{z}$ is the CGS containing the PO $\boldsymbol{\varphi}_\text{o}^{q}$ and a DC residual $\beta_l^{\text{r},q}=\big[  {\widehat{\boldsymbol{A}_{\text{d}}^{q}}^{+}}\cdot \boldsymbol{h}_{\text{s}}\big]_l$. The submatrix from the $L_{\text{d}}+1$ to $K$-th row of $\boldsymbol{H}^{\text{sep},q}$ is the projection of the static paths component into the dynamic paths' nullspace, $\boldsymbol{H}^{{\text{proj},q}}\!=\!{\widehat{\boldsymbol{A}_{\text{d,null}}^{q}}}^{\intercal}\!\cdot\!\boldsymbol{h}_{\text{s}}\!\cdot\!\boldsymbol{1}_{1\times T}+\boldsymbol{z}$, where the dynamic path components are eliminated by beam-nulling \cite{Multiple_Target_Doppler}. 

The PO is then estimated using $\boldsymbol{H}^{{\text{proj},q}}$. Applying the singular value decomposition, $\boldsymbol{H}^{{\text{proj},q}}\!=\!\boldsymbol{U}\boldsymbol{\Sigma}\boldsymbol{V}$, the differentiated PO is estimated as $\hat{\boldsymbol{\varphi}}_{\text{o}}^{q} \!\leftarrow\! \angle([\boldsymbol{V}^{\intercal}]_1)$. Next, the CGS estimate for each dynamic path is obtained by compensating for the PO, $\hat{\boldsymbol{\beta}}_{l}^{q} \!\leftarrow\! \text{diag}(\text{e}^{-\text{j}\hat{\boldsymbol{\varphi}}_\text{o}^{q}})\!\cdot\! \boldsymbol{\beta}_{l}^{\text{PO},q}$. The resulting CGS estimate, given by  $\hat{\boldsymbol{\beta}}_{l}^{q}\!=\!\text{e}^{\text{j}\varphi_{\text{o}}^{\text{r},q}}\!\cdot\!\big(\boldsymbol{\beta}_{l}^{q}\!+\!\beta_l^{\text{r},q}\!\cdot\!\boldsymbol{1}_{T\times 1}\big)\!+\!\boldsymbol{z}$, accounts for the true CGS value $\boldsymbol{\beta}_{l}^{q}$ after rotation by the PO residual $\varphi_{\text{o}}^{\text{r},q}$ and translation by the DC residual $\beta_l^{\text{r},q}$ on the complex plain. Since the PO residual and DC residual are constant across snapshots, the CGS estimate retains the dynamic pattern of the true CGS value, thus containing the target's Doppler information, which is primarily used by subsequent sensing applications \cite{Performance_Bound_Asyn}.

\subsection{Extension to MIMO-ISAC}
\label{sec:extension}
The framework can be extended to a MIMO-ISAC system, enabling joint AoA-delay estimation and improving accuracy. 

Consider the receiver equipped with $M$ antennas, the joint spatial-frequency response vector for a path with AoA parameter $\theta$ and delay $\tau$ is given by $\boldsymbol{a}^{\text{M}}(\theta, \tau)\in\mathbb{C}^{M\!K\times1}$, 
\vspace{-3pt}
\begin{equation}
	\boldsymbol{a}^{\text{M}}(\theta, \tau)=\mathrm{e}^{\mathrm{j}\boldsymbol{p}(\theta)}\otimes\mathrm{e}^{-\mathrm{j}2\pi\boldsymbol{f}\tau},
	\vspace{-3pt}
\end{equation}
where $\boldsymbol{p}(\theta) \in\mathbb{R}^{M\times1}$ represents the phase difference between antennas introduced by AoA $\theta$, which is determined by the array shape. Corresponding to \eqref{eq:CSI_measurements}, the CSI measurement for multiple-antenna receiver is given by $\hat{\boldsymbol{H}}^{\text{M},q}\!\in\!\mathbb{C}^{M\!K\!\times\! T}$,
\vspace{-3pt}
\begin{equation}
	\hat{\boldsymbol{H}}^{\text{M},q} = \big(\boldsymbol{A}^{\text{M},q} \cdot {\boldsymbol{B}^{q}}^{\intercal}\big) \odot \boldsymbol{\Phi}^{\text{M}}(\boldsymbol{\tau}^{q}_{\text{o}}, \boldsymbol{\varphi}^{q}_{\text{o}}) + \boldsymbol{z},
	\label{eq:CSI_measurements_M}
	\vspace{-3pt}
\end{equation}
where \mbox{$\!\!\!\!\boldsymbol{A}^{\text{M},q}\!\!=\!\big[\boldsymbol{h}_{\text{s}}^{\!\text{M}}\, \boldsymbol {A}^{\text{M},q}_{\text{d}}\big] \!\!\in\!\! \mathbb{C}^{M\!K\!\times\! (\!L_{\text{d}}+1\!)}$} \mbox{is the joint spatial-fre-} \mbox{quency response matrix, with \mbox{$\!\boldsymbol{A}^{\!\text{M},q}_{\text{d}}\!\!=\!\!\big[\!\boldsymbol{a}^{\!\!\text{ M}\!}(\theta^{q}_{1},\!\tau^{q}_{1}),\!..,\boldsymbol{a}^{\!\text{M}\!}(\theta^{q}_{\!L_{\text{d}}},\!\tau^{q}_{\!L_{\text{d}}} )\!\big]$}} denoting the joint space-frequency response submatrix corresponding to the dynamic paths. 
The clock asynchronous term for the multi-antenna receiver is given by 
$\boldsymbol{\Phi}^{\text{M}}(\boldsymbol{\tau}^{q}_{\text{o}}, \boldsymbol{\varphi}^{q}_{\text{o}})=\big[{\mathrm e}^{{\mathrm j}\varphi^{q}_{\text{o},1}}\boldsymbol{a}^{\text{M}}(0, \tau^{q}_{\text{o},1}),\ldots,{\mathrm e}^{\mathrm{j}\varphi^{q}_{\text{o},T}}\boldsymbol{a}^{\text{M}}(0,\tau^{q}_{\text{o},T})\big]$.

For TO alignment and compensation, the extension to the multi-antenna model can be simply regarded as the expansion of the response vector's length. 
In the algorithm, this extension can be implemented by replacing the CSI measurement and TO compensation terms in \multicite{eq:covariance, eq:TO_estmation_subspace, eq:snapshot_alignment, eq:TO_estmation_covariance, eq:h_ref} with their multi-antenna versions. 
Specifically, the CSI measurement terms $\hat{\boldsymbol{h}}_t^{q}$ and $\boldsymbol{H}^{\text{w}}$ are respectively replaced by the column of $\hat{\boldsymbol{H}}^{\text{M},q}$ and the previously aligned segment of $\hat{\boldsymbol{H}}^{\text{M},q}$, and the TO compensation term $\text{diag}\!\big(\boldsymbol{a}\!(\tau_{\text{o}}^{\text{r},q})\!\big)$ is substituted by $\text{diag}\!\big(\boldsymbol{a}^{\text{M}}\!(0,\tau_{\text{o}}^{\text{r},q})\!\big)$. 

For delay estimation, incorporating joint space-frequency domain array signals extends the approach to a 2-D MUSIC algorithm for joint AoA-delay estimation similar to \cite{JADE_MUSIC}, thereby enhancing accuracy. 
By replacing the response vector $\boldsymbol{a}(\tilde{\tau}\!)$ and \mbox{the projection matrices $\boldsymbol{P}_{\!\text{n},\boldsymbol{h}_{\text{s}}\!}$ and $\boldsymbol{P}_{\!\text{n}\!}$ in \eqref{eq:modified_MUSIC} with} $\boldsymbol{a}^{\!\text{M}}\!\big(\tilde{\theta},\tilde{\tau}\!\big)\!$ and the projection matrices calculated from the TO-compensated multi-antenna signals, respectively, the modified MUSIC spectrum $S_{\tau}(\tilde{\tau})$ is expanded to a 2-D spectrum $S_{\theta,\tau}(\tilde{\theta},\tilde{\tau})$. The AoA-delay pair for dynamic target paths, $(\hat{\theta}^{q}_l,\hat{\tau}^{q}_l), l \in [1 .. L{\text{d}}]$, are then estimated through peak searching in $S_{\theta,\tau}(\tilde{\theta},\tilde{\tau})$.

The PO compensation and CGS estimation procedure can be extended to the multi-antenna model by directly replacing $\widehat{\boldsymbol{A}_{\text{d}}^{q}}$ and $\boldsymbol{H}^{\text{C},q}$ in \eqref{eq:signal_sep} with $\widehat{\boldsymbol{A}_{\text{d}}^{\text{M},q}}\!=\!\big[\boldsymbol{a}^{\text{M}}(\hat{\theta}^{q}_1,\hat{\tau}^{q}_1),\!..,\boldsymbol{a}^{\text{M}}(\hat{\theta}^{q}_{L_{\text{d}}},\hat{\tau}^{q}_{L_{\text{d}}})\big]$ and the TO-compensated multi-antenna signals, respectively. This adaptation improves CGS estimation accuracy by leveraging the spatial diversity and angular resolution.

\section{Simulation Evaluation}

\begin{figure*}[t]
	\centering
	\vspace{-18pt}
	\begin{minipage}[t]{0.66\textwidth}
		\centering
		\includegraphics[width=11.9cm, height=6.4cm]{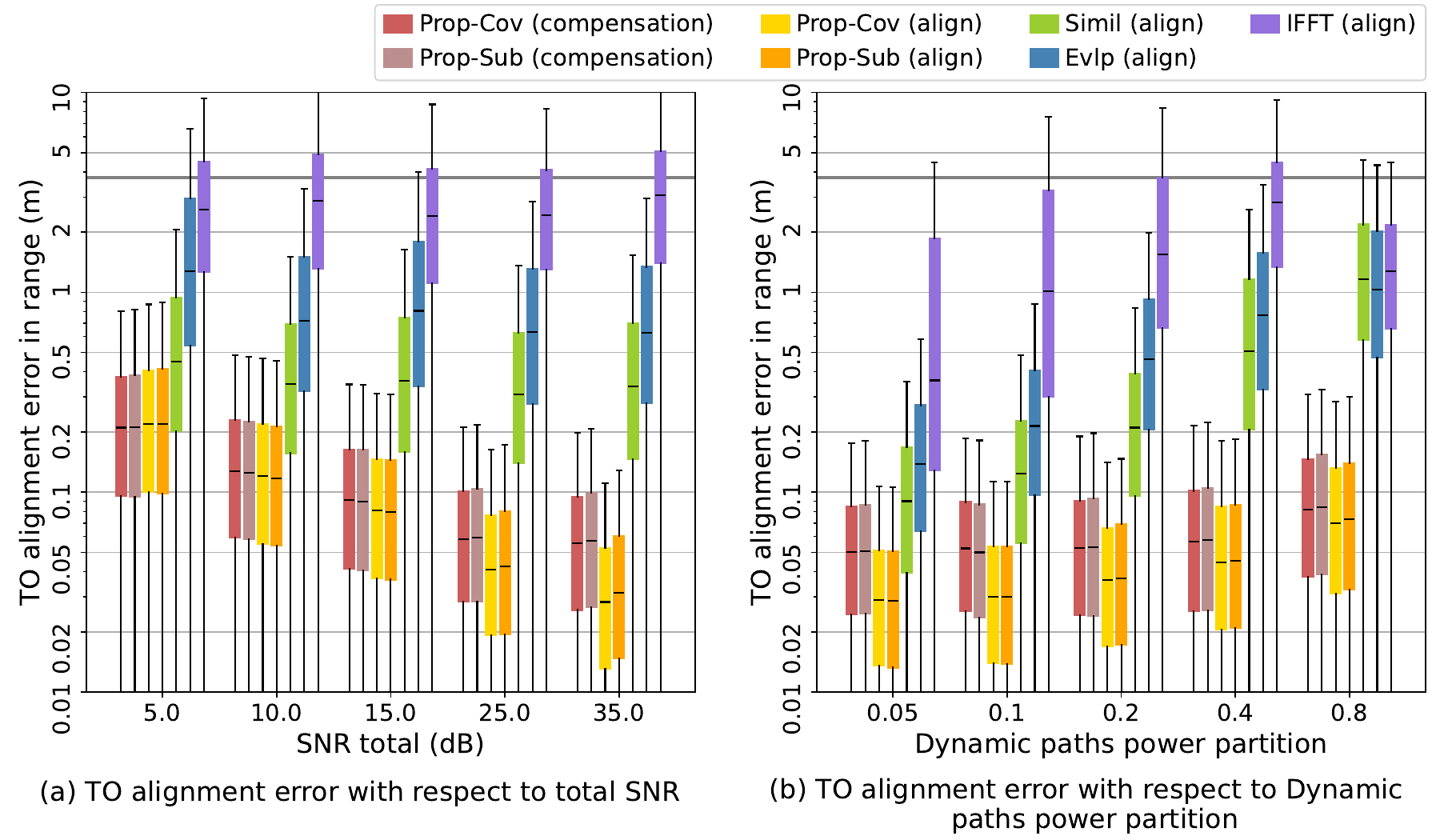}
		\vspace{-18pt}
		\captionsetup{margin={0pt, 4pt}}
		\caption{Comparison of TO alignment and compensation accuracy: (a) TO alignment/compensation error with respect to SNR, dynamic path power partition=$0.3$; (b) TO alignment/compensation error with respect to dynamic path power partition, SNR=$25\text{dB}$.}
		\label{fig:sim_TAE}
	\end{minipage}
	\hfill 
	\begin{minipage}[t]{0.33\textwidth}
		\centering
		\includegraphics[width=5.9cm, height=6.4cm]{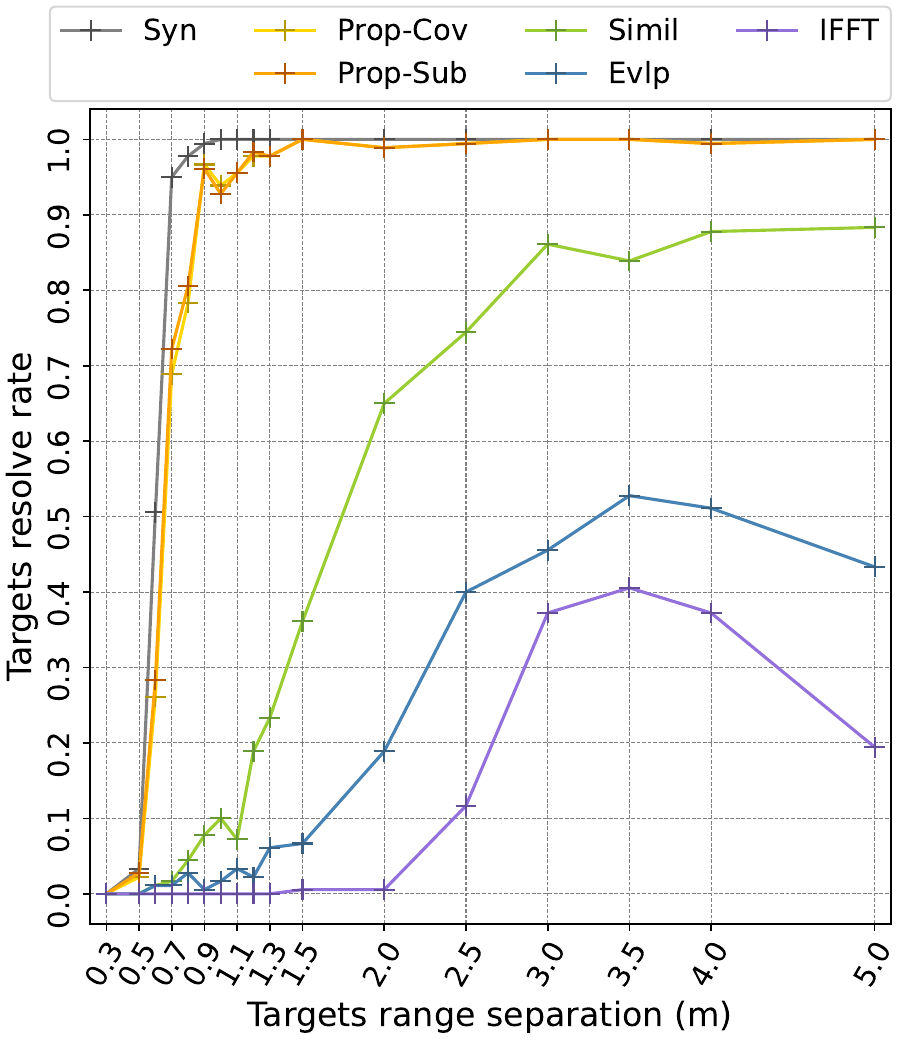} 
		\vspace{-18pt}
		\captionsetup{margin={4pt, 0pt}}
		\caption{Comparison of delay resolution, SNR=$25\text{dB}$, dynamic paths power partition=$0.3$.}
		\label{fig:sim_delay_resolution}
	\end{minipage}
	\vspace{-15pt}
\end{figure*}

\label{sec:evaluations}
In this section, we evaluate the performance of the proposed signal processing framework through numerical simulations.

The simulations consider a single-input single-output (SISO)-OFDM system employing 32 subcarriers ($K\!\!=\!\!32$) with a total bandwidth of $80\,\mathrm{MHz}$ ($\Delta f \!\!=\!\! 2.5 \,\mathrm{MHz}$). 
For each realization, the system performs 100 channel measurements ($T\!\!=\!\!100$) over a duration of $0.4\,\mathrm{s}$ ($\Delta T\!\!=\!\!4\,\mathrm{ms}$), yielding a CSI matrix $\boldsymbol{H} \!\!\in\!\! \mathbb{C}^{32 \!\times\! 100}$. The bi-directional CSI measurements for acquiring the reference static paths response $\boldsymbol{h}_{\text{s}}^{\text{ref}}$ has the length of $T_{\text{s}}\!\!=\!\!100$, and the measurement error for transmit/receive time, $\tau_t^{\text{BS},\text{T}}$, $\tau_t^{\text{BS},\text{R}}$, $\tau_t^{\text{UE},\text{T}}$, and $\tau_t^{\text{UE},\text{R}}$, are assumed to have a standard deviation of $2.5\,\text{ns}$. 

The simulations are conducted in randomly generated scenarios. 
Each scenario contains seven static paths and three target-reflected dynamic paths. 
The delays of the static paths are assumed to follow a Rayleigh distribution with a mean equivalent to a range of $12\,\mathrm{m}$, 
while the delays of the dynamic paths are uniformly distributed between $8$ to $20\,\mathrm{m}$, with a change rate between $0$ to $2\,\mathrm{m/s}$. 
The mean power of each path is inversely proportional to the square of its delay, and the CGS of the dynamic paths are assumed to be complex Gaussian distributed.

For each realization, we first apply the TO alignment algorithm described in Sec. \ref{sec:TO_Alignment} with a window size of $T_{\text{w}}\!=\!48$. 
The aligned snapshots, with $T\!=\!100$, are then used for TO residual compensation as detailed in Sec. \ref{sec:TO_residual_Compensation_A} and \ref{sec:TO_residual_compensation_B}.
Finally, the delay and CGS of each dynamic path are estimated using the methods presented in Sec. \ref{sec:delay_estimation} and \ref{sec:CGS_estimation}.

As baselines for comparison, we select several methods for asynchronous signal processing from existing research \cite{Optimal, Anchor, Spotfi, Sharp, JUMP}, adapting their core principles to our framework and experimental settings in an equivalent manner. 
The baselines include: 
(a) a similarity-based method, abbreviated as ``Simil'', which compensates for the TO and PO by maximizing the similarity between temporal snapshots in a recursive manner, based on the methods in \cite{Optimal, Anchor}; 
(b) an envelope-based method, abbreviated as ``Evlp'', which compensates for the TO by aligning the delay-domain envelopes across channel snapshots, following the principle of \cite{JUMP}; 
and (c) an inverse fast Fourier transform (IFFT)-based method, abbreviated as ``IFFT'', which compensates for the TO and PO by aligning the peaks in IFFT spectrum across snapshots, equivalent to the approaches in \cite{Spotfi, Sharp}. 
Additionally, the proposed covariance-based and subspace-based methods are abbreviated as ``Prop-Cov'' and ``Prop-Sub'' for short, respectively.
We analyze the impact of signal-to-noise ratio (SNR) and the power proportion of dynamic paths. The simulation results are summarized as follows.

\subsection{TO Alignment and Compensation}

The simulation results for TO alignment and compensation are illustrated in Fig. \ref{fig:sim_TAE}. Overall, the TO alignment accuracy of the proposed methods significantly outperforms the baselines. 

Comparing the TO alignment accuracy, as shown in Fig. \ref{fig:sim_TAE}a, the TO alignment error of proposed methods decreases with increasing SNR, achieving a centimeter-level median error at $15\,\text{dB}$, which is far less than the traditional resolution of 3.75 meters. 
In contrast, baseline methods achieve accuracies ranging from several decimeters to several meters, and their precision does not substantially improve with increasing SNR. 
This is because their references for TO compensation, essentially the delay domain profile or a dominant path peak, are biased as dynamic path gains vary. Such discrepancies are exacerbated in challenging scenarios with substantial dynamic path power proportions, NLoS conditions, and limited bandwidth.

\begin{figure*}[t]
	\centering
	\vspace{-18pt}
	\begin{minipage}[t]{0.66\textwidth}
		\centering
		\includegraphics[width=11.9cm, height=6.4cm]{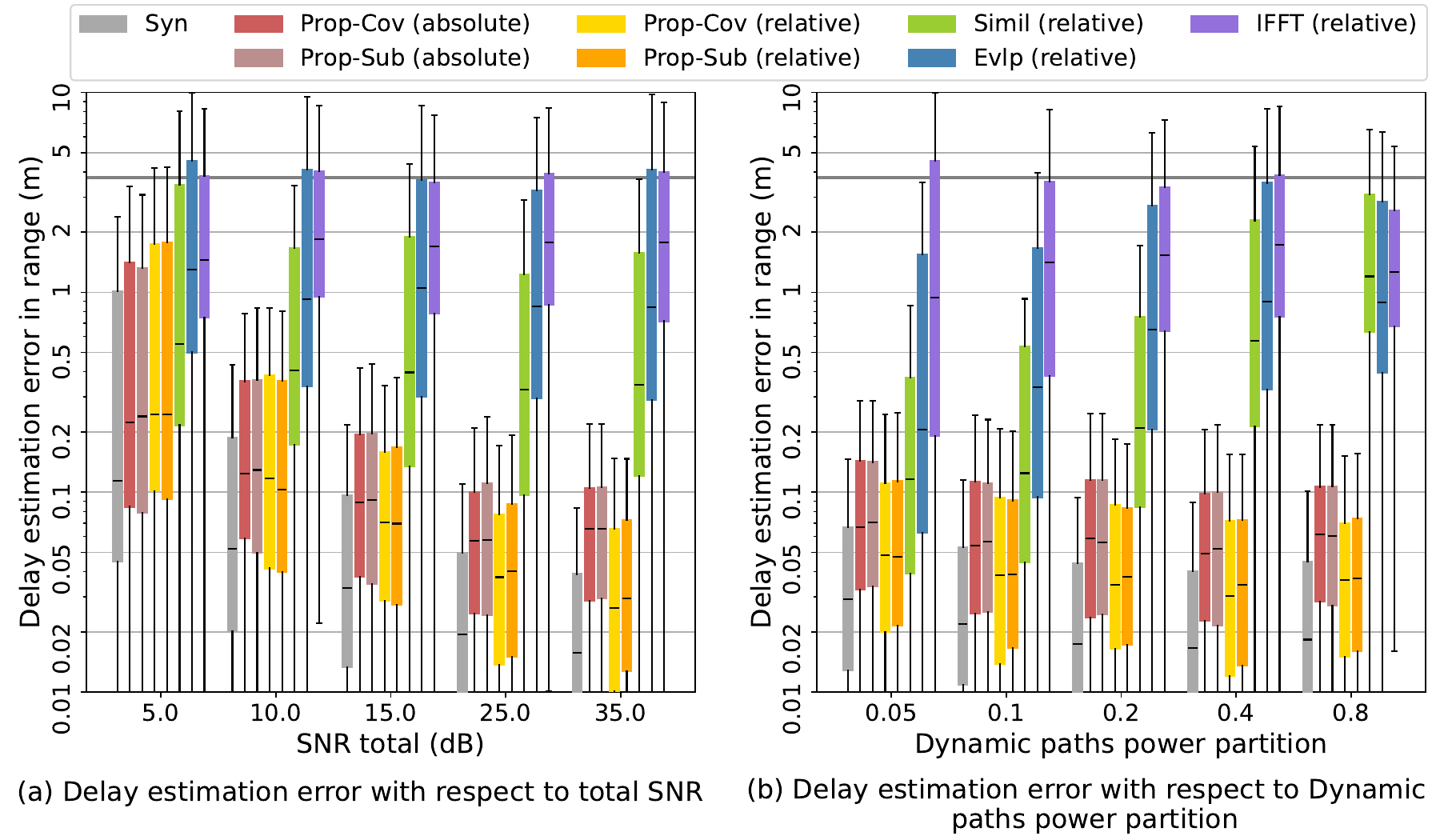}
		\vspace{-18pt}
		\captionsetup{margin={0pt, 4pt}}
		\caption{Comparison of target delay estimation accuracy: (a) delay estimation error with respect to SNR, dynamic path power partition=$0.3$; (b) delay estimation error with respect to dynamic path power partition, SNR=$25\text{dB}$.}
		\label{fig:sim_DEE}
	\end{minipage}
	\hfill 
	\begin{minipage}[t]{0.33\textwidth}
		\centering
		\includegraphics[width=5.9cm, height=6.4cm]{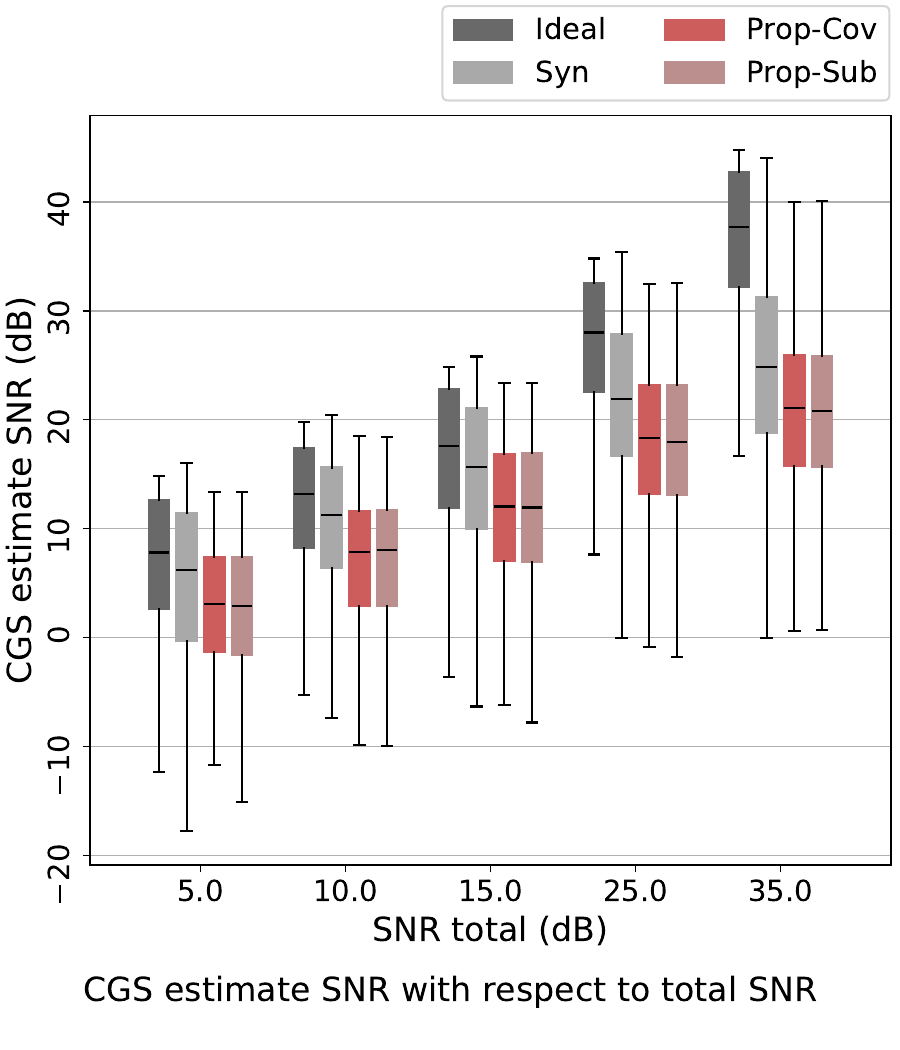} 
		\vspace{-18pt}
		\captionsetup{margin={4pt, 0pt}}
		\caption{CGS estimate SNR with respective to total SNR, dynamic paths power \mbox{proportion $=0.3$}.}
		\label{fig:sim_CGS}
	\end{minipage}
	\vspace{-12pt}
\end{figure*}

As illustrated in Fig. \ref{fig:sim_TAE}b, at moderate SNR of $25\,\text{dB}$, the TO alignment error of the proposed methods gradually increases with the rising proportion of dynamic path power proportion, yet remains a centimeter-level median accuracy. In contrast, the error of the baselines significantly increases as the dynamic path power proportion increases. Moreover, the proposed methods also outperform the baselines at low dynamic path power proportions, owing to their maximum likelihood design. 

Compared to the relative TO estimation accuracy, the absolute TO estimation introduces an additional error of a few centimeters, attributable to errors in the TO residual compensation procedure.

\subsection{Delay Estimation}

The delay estimation error for three-target scenarios is shown in Fig. \ref{fig:sim_DEE}. 
Similar to the results of TO alignment/compensation, the delay estimation error of the proposed methods shows a significant decrease as the SNR increases, reaching centimeter-level accuracy at $15\,\text{dB}$, and maintaining low error levels as the dynamic paths power proportion increases.
Compared to results with synchronized signals, the proposed algorithm shows small variations in relative delay estimation error that remains within $0$ to $3\,\text{dB}$ across different SNRs and dynamic path power proportions, demonstrating its effectiveness and robustness. Additionally, the absolute delay error shows a few centimeters increase compared to the relative one, which is negligible in practical scenarios. 
Notably, the proposed methods can still estimate the path delay with centimeter-level accuracy when the channel is dominated by dynamic paths (dynamic paths power proportion = 0.8). This suggests the strong potential of the proposed algorithm in applications involving moving transceivers.

The simulation results for delay resolution performance are summarized in Fig. \ref{fig:sim_delay_resolution}. 
This simulation considers estimating the delays of two targets separated by a delay difference of $\tau_{\text{sep}}$. The targets are regarded as successfully resolved when their relative delay estimation error is less than $\tau_{\text{sep}}/2$ \cite{Resolution_probability_2}. The metric, probability of resolution \cite{Resolution_probability_1}, is calculated as the ratio of successful resolutions to the total number of tests.
The traditional delay resolution of the setup is given by $\frac{C}{B_{\text{w}}}\!\!=\!\!3.75\,\text{m}$.
Using synchronous signals, the classic super-resolution MUSIC algorithm effectively resolves targets spaced much closer than the traditional delay resolution, achieving a $90\%$ resolution probability at a target separation of $0.7\,\text{m}$.
Using the asynchronous signal, our proposed methods achieve a probability of resolution of $90\%$ at a target separation of $0.9\,\text{m}$, with the probability-of-resolution curve closely approaching that of MUSIC using synchronous signals.
This demonstrates that our methods, through precise TO alignment, effectively enable super-resolution delay estimation, with resolution closely approaching that of synchronous systems. 
In contrast, the baselines, combining the MUSIC algorithm and the existing methods for addressing the clock asynchronism, only achieve mediocre probabilities of resolution near the traditional delay resolution. This is attributed to their lower TO alignment accuracy, which does not adequately support super-resolution.

\subsection{CGS Estimation}

\begin{figure*}[t]
	\vspace{-18pt}
	\centering
	\includegraphics[width=16.5cm]{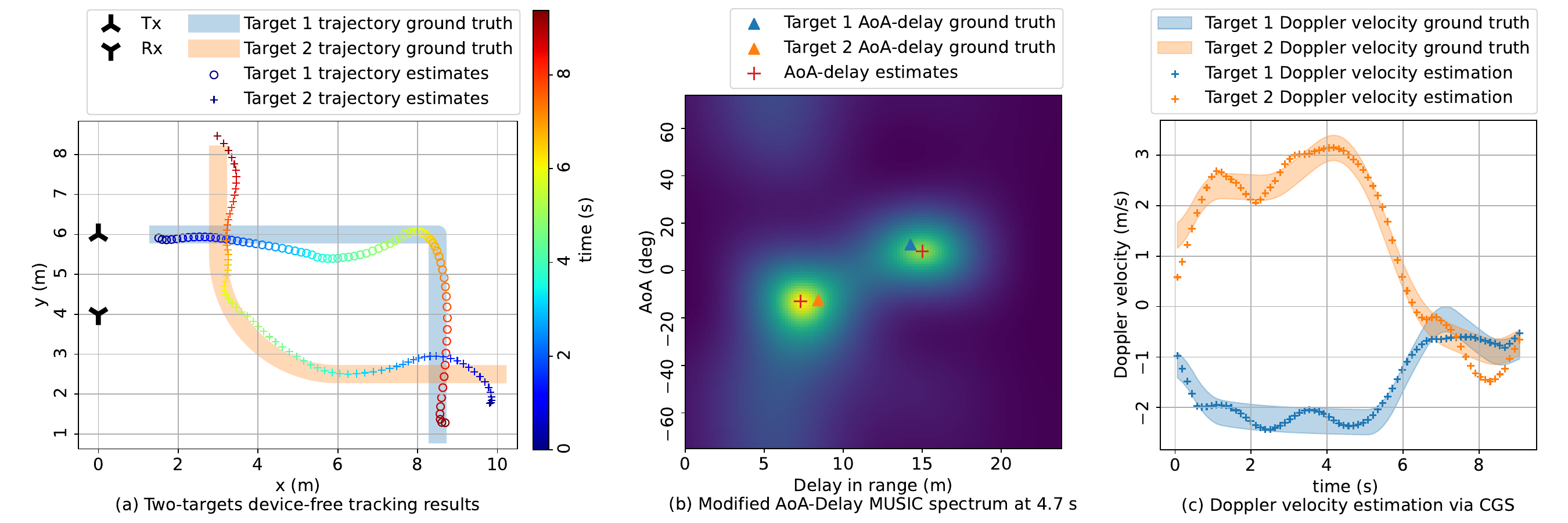}
	\vspace{-7pt}
	\caption{Practical device-free tracking results in a two-targets scenario: (a) Targets trajectory estimation; (b) Modified MUSIC spectrum for target paths AoA-Delay estimation at 4.7s; (c) Targets Doppler velocity estimation results using CGS parameters. }
	\label{fig:exp}
	\vspace{-12pt}
\end{figure*}

The SNR of the CGS estimates produced by the proposed framework is shown in Fig. \ref{fig:sim_CGS}. 
The SNR metric for the CGS estimates, denoted as $\gamma_{\beta}$, is defined as the ratio of the power of CGS ground truth to the power of the CGS estimation error\footnote{\setstretch{0.8}\noindent As detailed in \cite{Performance_Bound_Asyn}, the CGS estimate represents the ground truth's translation and rotation on the complex plane. Thus, the CGS estimation error, reflecting the accuracy of estimating the CGS's dynamic pattern, is calculated after aligning the CGS estimate to the ground truth via translation and rotation.}, $\gamma_{\beta}\!=\!\nicefrac{|\boldsymbol{\beta}_l^q|^2}{|\hat{\boldsymbol{\beta}}_l^q-\boldsymbol{\beta}_l^q|^2}$. This metric predicts the performance of subsequent sensing applications that rely on the CGS estimates. 
The ``Ideal'' CGS estimates SNR, calculated as $\gamma_{\beta}^{\text{Ideal}}\!=\!\nicefrac{K|\boldsymbol{\beta}_l^q|^2}{T\sigma^2}$, represents the idealized performance of CGS estimation that fully utilize the subcarrier diversity while ignoring clock asynchronism, delay estimation error, and coupling between dynamic paths. The distribution of $\gamma_{\beta}^{\text{Ideal}}$ is determined by the power variation across different dynamic paths. 
Compared to the ``Ideal'' cases, the CGS estimates SNR using synchronized signals decreases by $1$ to $12$ $\text{dB}$ due to coupling between dynamic paths and delay estimation errors. 
Using signals with clock asynchronism, the median CGS estimate SNR of our proposed methods is $3$ to $4$ $\text{dB}$ lower than that in the synchronized case. 
Note that this moderate gap is achieved in challenging scenarios for PO compensation, where a single-antenna receiver is used, and multiple dynamic and static paths concentrate within a narrow delay range. 
This demonstrates that the proposed framework supports effective CGS estimation for asynchronous ISAC systems, albeit with a reasonable performance degradation compared to the synchronized case, due to additional uncertainty associated with time-varying TO and PO.

\section{Use Case: Multi-Target Device-Free Tracking}
\label{sec:use_case}
To validate its practical effectiveness, we implement the proposed framework on a commercial Wi-Fi platform, achieving multi-target device-free tracking.

Specifically, the framework employs a single-antenna transmitter and a three-antenna receiver, both equipped with Intel 5300 NIC and the CSI tool \cite{CSItool} to measure the CSI. 
The transceivers operate on channel 102 at a $5.51\,\text{GHz}$ center frequency with a $40\,\text{MHz}$ bandwidth.
In a typical scenario and deployment for device-free tracking, where two human targets simultaneously walk along the trajectories as illustrated in Fig. \ref{fig:exp}a, the transceivers measure the CSI at a rate of 400 snapshots per second, and the proposed framework is applied to estimate AoA, delay, and CGS of the dynamic target paths. Subsequently, for each target, the Doppler velocity is estimated using CGS estimates through the method described in \cite{DopplerEstimation}, and the target trajectories are estimated via Kalman filtering employing the AoA, delay, and Doppler velocity estimates. 

Fig. \ref{fig:exp} illustrates the results of device-free tracking and intermediate parameter estimation, highlighting the practical effectiveness of the proposed framework in providing accurate parameter estimation for multiple target paths.
Specifically, Fig. \ref{fig:exp}a illustrates the trajectory estimation for two simultaneously tracked targets, achieving a root mean square error of 0.44 meters. 
Given that only a $40\,\text{MHz}$ bandwidth and three antennas are employed, and reflection points can vary along the human body by decimeters, this accuracy is reasonable and effective. 
Notably, intersections in the trajectories occur at points of AoA or delay. Since the framework employs a two-dimensional AoA-Delay joint estimation, it effectively resolves the targets, unless both AoAs and delays are identical across the targets.
As shown in Fig. \ref{fig:exp}b, the two-dimensional modified MUSIC spectrum for AoA-Delay exhibits the framework's capability to resolve the targets within the joint AoA-delay domain and provides accurate AoA and delay estimations for each target.
The results of the Doppler velocity estimates are presented in Fig. \ref{fig:exp}c. Despite the fluctuation in limb movements and walking speeds, the Doppler velocity estimations for the two targets are accurate, indicating that the framework provides reliable CGS estimates for each target.

\section{Conclusion}
\label{sec:conclusion}
In this paper, we introduce a subspace-based framework for bi-static sensing in challenging scenarios. This framework utilizes asynchronous CSI to generate AoA, delay, and CGS estimates for multiple dynamic target paths. 
Simulation and experiment results demonstrate that the framework significantly outperforms existing solutions in terms of parameter estimation accuracy and delay resolution. 
Notably, in the challenging single-antenna scenario with multiple closely-distributed targets, the proposed framework achieves performance that is remarkably close to that of a fully synchronized system, with only a 0 to 3 dB gap in delay estimation, a 3 to 4 dB gap in CGS estimation, and a minor discrepancy in the delay probability-of-resolution curve. Considering the challenging scenario applied in the tests, these gaps are modest and reasonable, laying a strong foundation for reliable fine-grained sensing applications.

We anticipate that this framework will play \mbox{a valuable role} in supporting and improving the performance of bi-static ISAC sensing applications while streamlining their development. 
Furthermore, the proposed algorithms hold promise for advancing bi-static sensing technologies involving moving transceivers, as they demonstrate effectiveness even in dynamic-dominant scenarios. 

\setstretch{0.956}
\bibliography{citations}
\setstretch{1.00}

\appendices
\vspace{-2pt}
\section{Derivation of Subspace-Based Relative TO Estimation}
\vspace{-1pt}
\label{Apd:TO_estimation_subspace}
Due to additive complex Gaussian noise, the log-likelihood function of $\hat{\boldsymbol{h}}_t^q$ is expressed as 
\vspace{-5pt}
\begin{equation}
	\setlength{\jot}{-2pt}
	\begin{aligned}
		&\!\!\ln\!\Big(f\big(\hat{\boldsymbol{h}}_t^q;\Delta\tau_{\!\text{o},t}^{q},\boldsymbol{A}^{\text{r},q},\boldsymbol{b}_{t}^{q}\!\!\cdot\! \mathrm{e}^{\mathrm{j}\varphi_{\!\text{o},t}^{q}}, {\sigma^{q}_{t}}^2\big)\!\Big)\!=
		-K\ln(\pi) \\
		&\,\, -\!K \!\ln\!\big({\sigma^{q}_{t}}^2\big) \!-\!\frac{1}{\,{\sigma^{q}_{t}}^2\!}\bigg|\hat{\boldsymbol{h}}_t^q \!\!-\! \text{diag}\!\Big(\!\boldsymbol{a}\big(\!\Delta\tau_{\text{o},t}^{q}\!\big)\!\!\Big) \!\!\cdot\!\! \boldsymbol{A}^{\text{r},q} \!\!\cdot\! {\boldsymbol{b}^{q}_t} \!\!\cdot\! \mathrm{e}^{\mathrm{j}\varphi_{\!\text{o},t}^{q}}\!\bigg|^2 \!\!,
	\end{aligned}
	\label{eq:log_likelihood_subspace}
	\vspace{-7pt}
\end{equation}
where $\Delta\tau_{\!\text{o},t}^{q}$ is the parameter of interest, $\boldsymbol{b}_{t}^{q}\!\!\cdot\! \mathrm{e}^{\mathrm{j}\varphi_{\!\text{o},t}^{q}}$, and ${\sigma^{q}_{t}}^2$ are nuisance parameters, and $\boldsymbol{A}^{\text{r},q}$ is implicitly a known parameter via the projection matrix $\boldsymbol{P}_{\mathrm{n}}^{\,\text{r},q}$.

The MLE is to find the joint maximum of \eqref{eq:log_likelihood_subspace} with respect to the parameters $\Delta\tau_{\!\text{o},t}^{q}$, $\boldsymbol{b}_{t}^{q}\!\!\cdot\! \mathrm{e}^{\mathrm{j}\varphi_{\!\text{o},t}^{q}}$, and ${\sigma^{q}_{t}}^2$. This joint maximum can be equivalently found solely with respective to $\Delta\tau_{\!\text{o},t}^{q}$, using the profile likelihood \cite{Concentrated_likelihood}.
Given the values of $\boldsymbol{A}^{\text{r},q}$ and the MLE of $\Delta\tau_{\text{o},t}^{q}$, denoted as $\widehat{\Delta\tau_{\text{o},t}^{q}}$, the MLE of $\boldsymbol{b}_{t}^{q}\!\!\cdot\! \mathrm{e}^{\mathrm{j}\varphi_{\!\text{o},t}^{q}}$ and ${\sigma^{q}_{t}}^2$ are straightforwardly given by 
\vspace{-5pt}
\begin{equation}
	\setlength{\jot}{-1pt}
	\begin{aligned}
		&\widehat{\boldsymbol{b}_{t}^{q}\!\!\cdot\! \mathrm{e}^{\mathrm{j}\varphi_{\!\text{o},t}^{q}}}=\big({\boldsymbol{A}^{\text{r},q}}^{\dagger}{\boldsymbol{A}^{\text{r},q}}\big)^{-1}{\boldsymbol{A}^{\text{r},q}}^{\dagger} \text{diag}\Big(\boldsymbol{a}^{*}\!\big(\widehat{\Delta\tau_{\text{o},t}^{q}}\big)\!\Big) \hat{\boldsymbol{h}}_t^q \\
		&\widehat{{{\sigma}^{q}_{t}}^2}=\frac{1}{K}\bigg|\boldsymbol{h}_t^q - \text{diag}\Big(\boldsymbol{a}\big(\widehat{\Delta\tau_{\text{o},t}^{q}}\big)\!\Big) \cdot {\boldsymbol{A}^{\text{r},q}} \cdot \widehat{\boldsymbol{b}_{t}^{q}\!\!\cdot\! \mathrm{e}^{\mathrm{j}\varphi_{\!\text{o},t}^{q}}} \bigg|^2 \!\!.
	\end{aligned}
	\label{eq:beta_sigma_MLE}
	\vspace{-6pt}
\end{equation}
Replacing ${\boldsymbol{b}_{t}^{q}\!\!\cdot\! \mathrm{e}^{\mathrm{j}\varphi_{\!\text{o},t}^{q}}}$ and ${{{\sigma}^{q}_{t}}^2}$ in \eqref{eq:log_likelihood_subspace} with their estimates in \eqref{eq:beta_sigma_MLE}, the profile likelihood function is given by 
\vspace{-5pt}
\begin{equation}
	\setlength{\jot}{-1pt}
	\!\!\!\!\!\!\begin{aligned}
	\!\!\!\!\!&\!\!\ln\Big(f\big(\hat{\boldsymbol{h}}_t^q;\Delta\tau_{\!\text{o},t}^{q}\big)\Big)=-K\ln(\pi \mathrm{e})-\ln\Big(\frac{1}{K}F(\Delta\tau_{\!\text{o},t}^{q})\!\Big), \\
	&\!\!\text{where}\, 
		F\!\big(\!\Delta\tau_{\!\text{o},t}^{q}\big)\!\!=\!{\hat{\boldsymbol{h}}_t^q \,\!}^{\dagger}\!\!\!\cdot\!\text{diag}\!\big(\!\boldsymbol{a}({\!\Delta\tau_{\text{o},t}^{q}})\!\big) \!\!\cdot\! \boldsymbol{P}_{\mathrm{n}}^{\,\text{r},q} \!\!\cdot\! \text{diag}\!\big(\!\boldsymbol{a}^{\!*}\!({\!\Delta\tau_{\text{o},t}^{q}})\!\big) \!\!\cdot\! {\hat{\boldsymbol{h}}_t^q}.\!\!\!\!\!\!\!\!
	\end{aligned}
	\label{eq:cost_f_subspace}
	\vspace{-3pt}
\end{equation}
Thus, the MLE of $\Delta\tau_{\!\text{o},t}^{q}$ can be found by minimizing \eqref{eq:cost_f_subspace}, which leads to \eqref{eq:TO_estmation_subspace}.

\vspace{-4pt}
\section{Derivation of Covariance-Based Relative TO Estimation}
\vspace{-2pt}
\label{Apd:TO_estimation_covariance}
Due to the assumption of zero-mean Gaussian distributed path gains $\boldsymbol{b}_{t}^{q}\!\!\cdot\! \mathrm{e}^{\mathrm{j}\varphi_{\!\text{o},t}^{q}}$, it can be inferred from \eqref{eq:snapshot} that the snapshot $\boldsymbol{h}_t^q$ follows a Gaussian distribution,
\vspace{-5pt}
\begin{equation}
	\setlength{\jot}{1pt}
	\begin{aligned}
		&\boldsymbol{h}_t^q \sim \mathcal{CN}\!\Big(\!\boldsymbol{0}, \text{diag}\!\big(\!\boldsymbol{a}(\Delta\tau_{\text{o},t}^{q})\!\big)\boldsymbol{R}_{\mathrm{h}}^{\,\text{r},q} \text{diag}\!\big(\!\boldsymbol{a}^{\!*}\!(\Delta\tau_{\text{o},t}^{q})\!\big)\!\!\Big)\!, \,\\
		&\text{where} \quad
		\boldsymbol{R}_{\mathrm{h}}^{\,\text{r},q}\!=\!  \boldsymbol{A}^{\text{r},q}  \boldsymbol{R}_{\mathrm{b}}^{q}  {\boldsymbol{A}^{\text{r},q}}^{\dagger}   \!+\! \sigma^2 \!\boldsymbol{I}. \!\!\!\!
	\end{aligned}
	\label{eq:snapshot_distribution}
	\vspace{-3pt}
\end{equation}
Thus, the log-likelihood of the snapshot can be written as 
\vspace{-5pt}
\begin{equation}
	\setlength{\jot}{1pt}
	\begin{aligned}
		&\!\!\ln\!\!\Big(\!p\big(\hat{\boldsymbol{h}}_t^q ; \Delta\tau_{\!\text{o},t}^{q}, \boldsymbol{R}_{\mathrm{h}}^{\,\text{r},q}\big)\!\Big)\!=
		-\!K\ln(\pi) -\!\ln\!\!\Big(\!\!\det\!\!\big(\boldsymbol{R}_{\mathrm{h}}^{\,\text{r},q}\big)\!\Big) \\
		&\quad\quad\quad\quad - \!{\hat{\boldsymbol{h}}_t^q\,\!}^{\dagger} \!\text{diag}\!\Big(\!\boldsymbol{a}\big(\!\Delta\tau_{\text{o},t}^{q}\!\big)\!\Big) {\boldsymbol{R}_{\mathrm{h}}^{\,\text{r},q}}^{-\!1} \!\text{diag}\!\Big(\!\boldsymbol{a}^{\!*}\!\big(\!\Delta\tau_{\text{o},t}^{q}\!\big)\!\Big) \hat{\boldsymbol{h}}_t^q .
	\end{aligned}
	\label{eq:log_likelihood_covariance}
	\vspace{-3pt}
\end{equation}
Although the covariance matrix ${\boldsymbol{R}_{\mathrm{h}}^{\,\text{r},q}}$ can be simply estimated as the covariance matrix of the previously aligned snapshots, $\boldsymbol{R}_{\text{w}}$, this estimate may be less accurate with a small snapshot number $T_{\text{w}}$ or a low SNR. Thus, we add a regulation term, $\widehat{\sigma^2}\boldsymbol{I}$, leading to an adjusted covariance matrix estimate $\widehat{\boldsymbol{R}_{\mathrm{h}}^{\,\text{r},q}} = \boldsymbol{R}_{\text{w}}+\widehat{\sigma^2}\boldsymbol{I}$. Substituting ${\boldsymbol{R}_{\mathrm{h}}^{\,\text{r},q}}$ with $\widehat{\boldsymbol{R}_{\mathrm{h}}^{\,\text{r},q}}$, the relative TO $\Delta \tau_{\text{o},t}^{q}$ can be estimated by maximizing \eqref{eq:log_likelihood_covariance}, which leads to \eqref{eq:TO_estmation_covariance}.

\end{document}